\documentclass[useAMS,usenatbib]{mn2e}
\usepackage{graphicx}
\usepackage{color}

\usepackage{longtable}
\usepackage{lscape}
\usepackage{subfigure}

\title[Modelling the ISO high resolution far-IR CO lines from Orion KL]
{Physical parameters for Orion KL from modelling its ISO
high resolution far-IR CO line spectrum}
\author[M. R. Lerate et al.]
{M. R. Lerate,$^{1,2}$, J. Yates$^1$, S. Viti$^1$, M. J.
Barlow$^1$, B. M. Swinyard$^2$,
\newauthor
G. J. White$^{2,3}$, J. Cernicharo$^4$
and J. R. Goicoechea$^5$\\
  $^1$University College London, Gower Street, London WC1E 6BT,
  U.K\\
  $^2$Rutherford Appleton Laboratory, Chilton, Didcot OX11 0QX,
  U.K \\
  $^3$ Department of Physics \& Astronomy, The Open University,
  Milton Keynes MK6 7AA \\
$^4$CSIC, Instituto de Estructura de la Materia, Serrano 121, 28006 Madrid, Spain.\\
 $^5$LERMA, UMR 8112, CNRS, Observatoire de Paris and Ecole Normale Superieure, 24 Rue Lhomond 75231 Paris Cedex 05,
 France}

\date{Released 2005 Xxxxx XX}

\pagerange{\pageref{firstpage}--\pageref{lastpage}} \pubyear{2005}

\def\LaTeX{L\kern-.36em\raise.3ex\hbox{a}\kern-.15em
    T\kern-.1667em\lower.7ex\hbox{E}\kern-.125emX}

\begin{document}

\label{firstpage}

\maketitle

   \begin{abstract}
  {As part of the first high resolution far-IR spectral survey of the
  Orion KL
  region (Lerate et al. 2006), we observed 20 CO
  emission lines with J$_{up}$=16 to J$_{up}$=39 (upper levels
  from $\approx$ 752 K to 4294 K above the ground state). Observations
  were taken using the
  Long Wavelength Spectrometer (LWS) on board the {\em Infrared
  Space Observatory (ISO)}, in its high resolution Fabry-P\'erot
  (FP) mode ($\approx$ 33 km s$^{-1}$). We present here
  an analysis of the final calibrated CO data, performed
  with a more sophisticated modelling technique than hitherto, including a
detailed analysis of the chemistry, and discuss similarities and
differences with previous results. The inclusion of chemical
modelling implies that atomic and molecular abundances are
time-predicted by the chemistry. This provides one of the main
differences with previous studies in which chemical abundances
needed to be assumed as initial condition.
  The chemistry of
the region is studied by simulating the
  conditions of the different known components of the KL region: chemical
  models for a
  hot core, a plateau and a ridge are
  coupled with an accelerated $\Lambda$-iteration (ALI) radiative
transfer model to predict line
  fluxes and profiles. We conclude that the CO transitions with
18$<$J$_{up}<$25 mainly arise from a hot core of diameter 0.02 pc
and a density of 10$^{7}$ cm$^{-3}$ rather from the plateau as
previous studies had indicated. The rest of the transitions
originate from shocked gas in a region of diameter $\approx$ 0.06
pc with densities ranging from
3$\times$10$^{5}$--1$\times$10$^{6}$ cm$^{-3}$. The resulting CO
fractional abundances are in the range
$X$(CO)=(7.0--4.7)$\times$10$^{-5}$. A high temperature post-shock
region at more than 1000 K is necessary to
  reach transitions with J$_{up}>$32, whilst transitions with J$_{up}<$18
probably originate from
  the extended warm component.
   Finally, we discuss the spatial origin of the CO emission compared
  with that of the next  most abundant species detected by the far-IR
survey
towards Orion KL: H$_{2}$O and OH.}
 \end{abstract}


\section{Introduction}
\begin{figure}
    \centering
   \includegraphics[width=8cm,height=8cm]{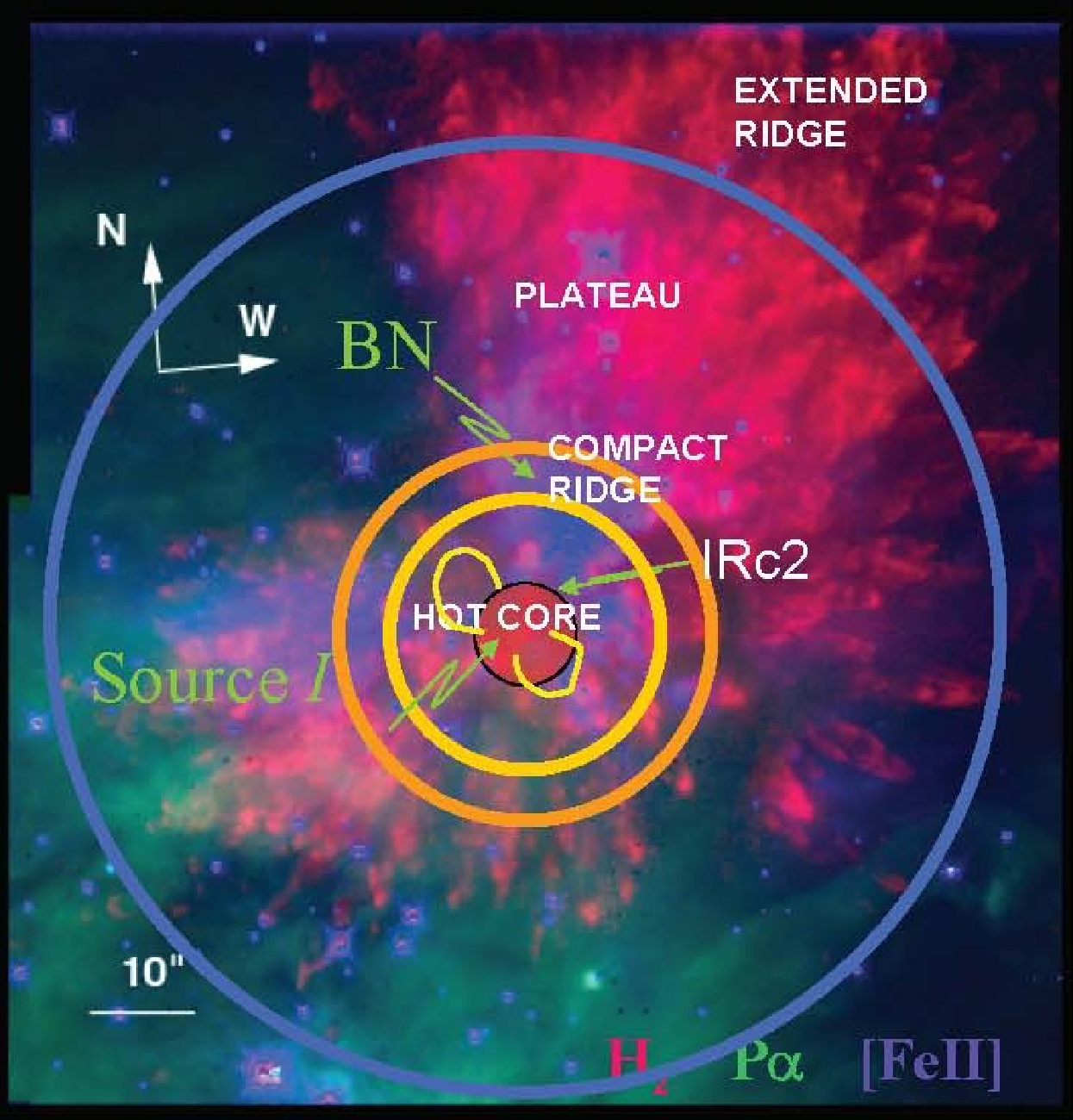}

    \caption{Sketch of the different components in Orion-KL over-plotted on the image of Schultz et al. (1999).
    Colours correspond to the emission in H$_{2}$ (red) , P$_{\alpha}$ (green) and [FeII] (blue). Note that the large circle is the approximate
    size of the LWS beam.}
        \label{kl_esque}
   \end{figure}

At a distance of 450 pc, the Orion molecular cloud is the nearest
massive star forming region. Molecular emission mainly comes from
the Orion Molecular cloud 1 (OMC1) which contains a number of
infrared-emitting regions such as the Kleinmann-Low (KL) region
(see the review by Genzel \& Stutzki, 1989). Since the first
spectroscopic measurements in the millimeter, submillimeter and
infrared (Storey et al., 1981; Genzel \& Stutzki, 1989; Sempere et
al., 2000), molecular emission has been shown to arise from
several physically distinct regions (Irvine, Goldsmith and Hjalmarson 1987):
the hot core, the compact
ridge, the ridge, the plateau and the PDR surrounding the
quiescent gas (see Figure~\ref{kl_esque}). The hot core and the
plateau are characterized by elevated temperature due to star
forming activity (the hot core) or as consequence of outflows and shocks
(the plateau), while the ridge represents extended, cooler, quiescent
material. \\

The powerful molecular outflow associated with OMC1 was discovered
by Zuckerman, Kuiper \& Rodriguez-Kuiper (1976), being later noted
as containing a weak bipolarity by Erickson et al. (1982), and
subsequently mapped with greater sensitivity and resolution.
Recent observations have shown that the observed bipolar high
velocity outflow probably originates at the position of radio
source I (Menten \& Reid, 1995; Gezari, Backman \& Werner, 1998;
Chandler \& Greenhill, 2002). Source I is just 0.5 arcsec from
IRc2, which had been believed to be the major source of the
luminosity in the region (L$\approx$ 10$^{5}L\odot$: Genzel \&
Stutzki, 1989). The hot core is probably a dense clump, or a set
of clumps, in the cavity surrounding these locations, with a gas
kinetic temperature near 200 K. The main heating mechanism is
radiative, warming up the grains
and thus inducing thermal evaporation of their icy mantles (Goldsmith et
al. 1983).\\

The Plateau emission comes from a mixture of outflows, shocks and
interaction regions where shock chemistry may be dominant, being
responsible for the large observed abundance of species such as
SO, SO$_{2}$ and SiO. These species can be easily formed in the
heated gas due to enhancements of the S, Si abundances following
grain disruption in the shocked region and OH abundances through
gas phase reactions at high temperature. However, fragile
molecules such as H$_{2}$CO or HDO have been also detected in the
plateau, suggesting that small dense condensations may be present
in the surrounding outflow (Blake et al. 1986). It is therefore
difficult to discriminate between processes due to
high-temperature gas-phase reactions or grain surface chemistry.\\

Interactions between the powerful outflow and quiescent ambient
material are thought to give rise to the compact ridge emission,
which is also referred to as the southern condensation. This
emission region appears to be spatially compact, being revealed in
observations of [C~{\sc i}] as a double-peaked structure from a shell
of radius $\approx$ 20 arcsec (Pardo et al., 2005). Finally, there
is a surrounding layer of warm photodissociated material, arising
from the non-ionising ultraviolet (UV) radiation from the
Trapezium OB stars (Rodriguez-Franco et al. 2001).\\

Among molecular tracers, CO is of particular interest, because of its
high abundance and
 because observations of different transitions can trace regions with
significant differences in their physical properties. CO transitions with
low J$_{up}$ values trace the less dense gas, with moderate temperatures
and constituting the bulk of the outflow, while the highest J$_{up}$
transitions trace either a dense or high temperature gas. \\

The first Orion-KL far-IR CO line detection was by Watson et al. (1980),
with the Kuiper Airborne Observatory (KAO). Their detections of
the J=21--20 and J=22--21 transitions, at 124 $\mu$m and 119
$\mu$m, indicated the presence of at least two different
temperature components, with temperatures between 400-2000 K and
a density in the range of 1-5$\times$10$^{6}$~cm$^{-3}$. In a more
recent analysis of ISO LWS data, Sempere et al. (2000) concluded
that the CO emission could be described by three temperature
components in the plateau and ridge, with
temperatures ranging from 300 to 2000 K and column densities of
N(CO)=10$^{17}$--10$^{19}$ cm$^{-2}$. \\
On the other hand, our far-IR survey of Orion-KL (Lerate et al.
2006) showed the presence of P-Cygni line profiles in H$_{2}$O and
OH transitions, revealing their outflow origin. Recent analyses of
these transitions (Goicoechea et al. 2006; Cernicharo et al. 2006)
inferred a beam averaged gas temperature of $\approx$ 100-150 K;
however, it is still not clear if these molecules form via high temperature
neutral-neutral reactions due to shocks or via desorption from
grains. The complexity of the region and the large beam size of
the ISO Long Wavelength Spectrometer($\approx$ 80 arcsec)
complicates the spatial differentiation of
the molecular emission.\\

In this paper we report our analysis of
the final ISO LWS-FP CO line detections from our recently
published far-IR spectral survey of Orion-KL.
Observations of multiple CO lines with the same
instrument are of great value, since all the
lines have the same instrumental calibration and the same
pointing. We used a new
approach to modelling the CO emission, in which chemical
simulations are carried out independently for each of the main
known physical components in the region. The resulting chemical
outputs are then coupled with non-local accelerated
$\Lambda$-iteration (ALI) radiative transfer modelling in order to
reproduce the observed line shapes and fluxes.

        \label{500_1000}


\section{Observations and Data reduction}

The dataset consists of 26 individual ISO observations in the LWS
high resolution modes L03 and L04. The complete list of Target
Dedicated Time identification numbers (TDT's) and detailed
description of the data processing can be found in Lerate et al.
(2006). The angular and spectral resolution achieved across the
LWS range was $\approx$ 80 arcsec and 33 km s$^{-1}$, although
this could vary slightly depending on the LWS detector (Gry et al.
2003). Table 1 summarises the detectors in which the different CO
J transitions are observed and the effective aperture for this
wavelength in arcsec. The last three columns are the instrumental
resolving power, the FWHM of the resolution element and the
overall calibration accuracy in $\mu$m. (Swinyard et al 1996). The
velocity accuracy achieved for the LWS FPs during the mission is
quoted as 6 km s$^{-1}$ for FPS and 13 km s$^{-1}$ for FPL (Gry et
al 2003). However, it should be noted that this is only for the
absolute velocity, i.e. comparing to an absolute standard of rest
for an individual object, and arises primarily in the uncertainty
in the knowledge of the absolute velocities of the sources used
for the calibration of the instrument. For comparison of relative
velocities of components within a given source, the better measure
to take is the statistical error on the centroid found from
repeated measurement of the same line on the same source over the
course of the mission (ibid). For both FPs we take this as $\pm$ 5
km s$^{-1}$ over all wavelengths.

Observations were centred at either RA J2000 05$^{\rm h}$ 35$^{\rm
m}$ 14.12$^{\rm s}$ Dec(2000)-- 05$^{\circ}$ 22$^{\prime}$
22.9$^{\prime\prime}$ on at RA(J2000) 05$^{\rm h}$ 35$^{\rm m}$
14.45$^{\rm s}$ Dec(J2000)-- 05$^{\circ}$ 22$^{\prime}$
30.0$^{\prime\prime}$, including both the BN object and IRc2 in
the beam.
\begin{table*}
\centering
\begin{tabular}{ c c c c c c c}
\hline\hline  J range & FP & Detector& Effective aperture
(arcsecs) & Resolving power & Resolution element ($\mu$m) & RMS accuracy\\
\hline
15--13 & FPL&LW5 & 66.4 & 8500 & 0.0209& 2.7$\times$10$^{-3}$\\
18--16 & FPL&LW4 & 69.4 & 8900 & 0.0180& 2.7$\times$10$^{-3}$\\
21--19 & FPL&LW3 & 71.0 & 9250 & 0.0153& 2.7$\times$10$^{-3}$\\
25--21 &FPL &LW2 & 77.8 & 9600 & 0.0127& 2.7$\times$10$^{-3}$\\
30--24 & FPL&LW1 & 77.2 & 9700 & 0.0105& 2.7$\times$10$^{-3}$\\
34--29 & FPL&SW5 & 79.0 & 9200 & 0.0092& 2.7$\times$10$^{-3}$\\
39--33 &FPL &SW4 & 81.8 & 7800 & 0.0097 & 2.7$\times$10$^{-3}$\\
45--38 & FPS&SW3 & 87.0 & 8200 & 0.0081& 8$\times$10$^{-4}$\\
52--45 & FPS & SW2 & 84.6 & 8450 & 0.0066 & 8$\times$10$^{-4}$\\
 \hline
 \end{tabular}
 \caption{J range of observed CO transitions for each detector and their effective radius aperture, resolving power
 and resolution element.}
 \label{co_detects}
 \end{table*}

The data processing was carried out using Offline Processing
(OLP), the LWS interactive Analysis (LIA) package version 10 and
the ISO Spectral Analysis Package (ISAP). Further processing was
also carried out by including dark current optimisation,
deglitching and removal of the LWS grating residual profile
(Lerate et al. 2006). A full grating spectrum with lower spectral
resolution was also acquired but these data suffer from detector
saturation effects. Thus only the Fabry-P\'erot observations are
considered in the present work.

\section{The chemical and radiative transfer models}

The three main components in Orion-KL (hot
core, plateau and the ridge) were chemically modelled.
The model is made from a two stage calculation: in Stage I
the chemical evolution from diffuse gas to a dense core is
computed. In Stage II
the chemical evolution of the plateau and ridge are investigated. \\

The time-dependent outputs from the chemical model are the
fractional abundances of gas-phase and grain surface species. The
fractional abundance of CO was then coupled with a radiative
transfer model and line profiles and intensities directly compared
with observations. The method and the models are similar to those
used by by Benedettini et al. (2006) to study the chemical
structure of molecular clumps along chemically rich outflows. Each
component is modelled independently. The final synthetic line
spectrum is obtained by adding up the contribution of the three
components.

\subsection{The chemical models: the hot core, the plateau and the ridge}

The hot core model is described by Viti et al. (2004a). It is a
modification of the model described by Viti \& Williams (1999),
where more detailed information about the molecules included and
main reactions can be found. The model follows the chemical
evolution of a free-fall collapsing cloud. The chemical evolution
of the molecular species is then investigated after typical
densities of hot cores ($\sim$ 10$^7$ cm$^{-3}$) have been
reached. The effect of an infrared source is simulated by an
increase in the gas and dust temperature (see Viti et al. 2004a
for more details). The initial and final densities of the
collapsing core are 4$\times$10$^{2}$~cm$^{-3}$ and 10$^{7}$
cm$^{-3}$. The diameter of the core is 10 $^{\prime\prime}$
($\approx$ 0.02 pc), as estimated by observations of hot cores
(e.g. Blake, Sutton and Masson, 1986) and the final visual
extinction is $\approx$ 400 mag. During collapse, gas-phase
species accrete, or freeze-out, onto the dust. The efficiency of
accretion, f$_{r}$, is a free parameter and a grid of conditions
for different f$_{r}$ values was investigated, from
f$_{r}$=0.03--0.63 ( f$_{r}$=1 means 100\% efficiency). The best
fit model was for f$_{r}$=0.23,
which led to 98\% of the CO in the form of ice at the end of Stage I.\\

\begin{figure}
    \centering
    \includegraphics[width=8cm,height=8cm]{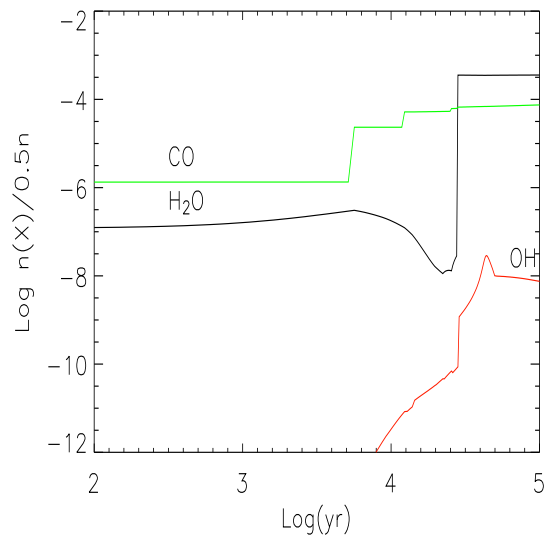}
    \caption{Time evolution of the abundances of CO, OH and H$_{2}$O in the Hot Core
    model.}
   \label{HC}
   \end{figure}

The models used to simulate the plateau and ridge are described by
Viti et al. (2004b). As in the hot core model, we have a two phase
calculation, where the gravitational collapse occurs in Stage I.
Both the ridge and the plateau components are believed to be
affected by shocks due to the outflows activities. In Stage II we
therefore simulate the presence of non-dissociative shocks by an
increase in temperature at an age of $\approx$ 1000 yr,
corresponding to the dynamical timescale of the main outflow
observed in the KL region (Cernicharo et al. 2006). The efficiency
of freeze-out during Stage I, the shock temperatures, and the
unshocked gas temperatures are free parameters and
Table~\ref{plat_para} lists the choice of these parameters for the
grid investigated. The parameters were chosen based on the results
of previous studies of the Orion-KL components (Blake et al. 1987,
Genzel and Stutzki 1989).

\begin{table*}
\centering
\begin{tabular}{ c c c c c c c c c c}
\hline\hline  Model &Density (cm$^{-3}$) & T$_{shock}$ (K)&
T$_{gas}$ (K) &mco \%
& Model &Density (cm$^{-3}$) & T$_{shock}$ (K)& T$_{gas}$ (K) & mco \% \\
\hline

PL1&3$\times$10$^{5}$ & 200 & 80 & 40 \vline & PL9&1$\times$10$^{6}$ & 300 & 90 & 60 \\
PL2&3$\times$10$^{5}$ & 300 & 90 & 60 \vline &  PL10&1$\times$10$^{6}$ & 300 & 90 & 80 \\
PL3&3$\times$10$^{5}$ & 500 & 90 & 60 \vline & PL11&1$\times$10$^{6}$ & 500 & 90 & 60\\
PL4&3$\times$10$^{5}$ & 500 & 90 & 80 \vline & PL12&1$\times$10$^{6}$ & 500 & 90 & 80 \\
PL5&3$\times$10$^{5}$ & 1000 & 100 & 60 \vline & PL13&1$\times$10$^{6}$ & 1000 & 100 &60\\
PL6&3$\times$10$^{5}$ & 1000 & 100 & 80 \vline & PL14&1$\times$10$^{6}$ & 1000 & 100 &80\\
PL7&3$\times$10$^{5}$ & 2000 & 100 & 60 \vline & PL15&1$\times$10$^{6}$ & 2000 & 100 &60\\
PL8&3$\times$10$^{5}$ & 2000 & 100 & 80 \vline & PL16&1$\times$10$^{6}$ & 2000 & 100 &80\\
\hline
RG1&1$\times$10$^{4}$ & no shock & 70 & 40 \vline & RG3&5$\times$10$^{4}$ & no shock & 80 & 40\\
RG2&1$\times$10$^{4}$ & no shock & 70 & 20 \vline & RG4&5$\times$10$^{4}$ & no shock & 80 & 20\\

 \hline
 \end{tabular}
 \caption{List of Plateau and Ridge models and their parameters: density, maximum temperature reached by the gas in the shock simulation, gas temperature after cooling,
 and  the percentage of mantle CO (mco) given by the freeze-out parameter
 at the end of Stage I of the chemical model.}
 \label{plat_para}
 \end{table*}

  \begin{figure}
    \centering
\includegraphics[width=8cm,height=8cm]{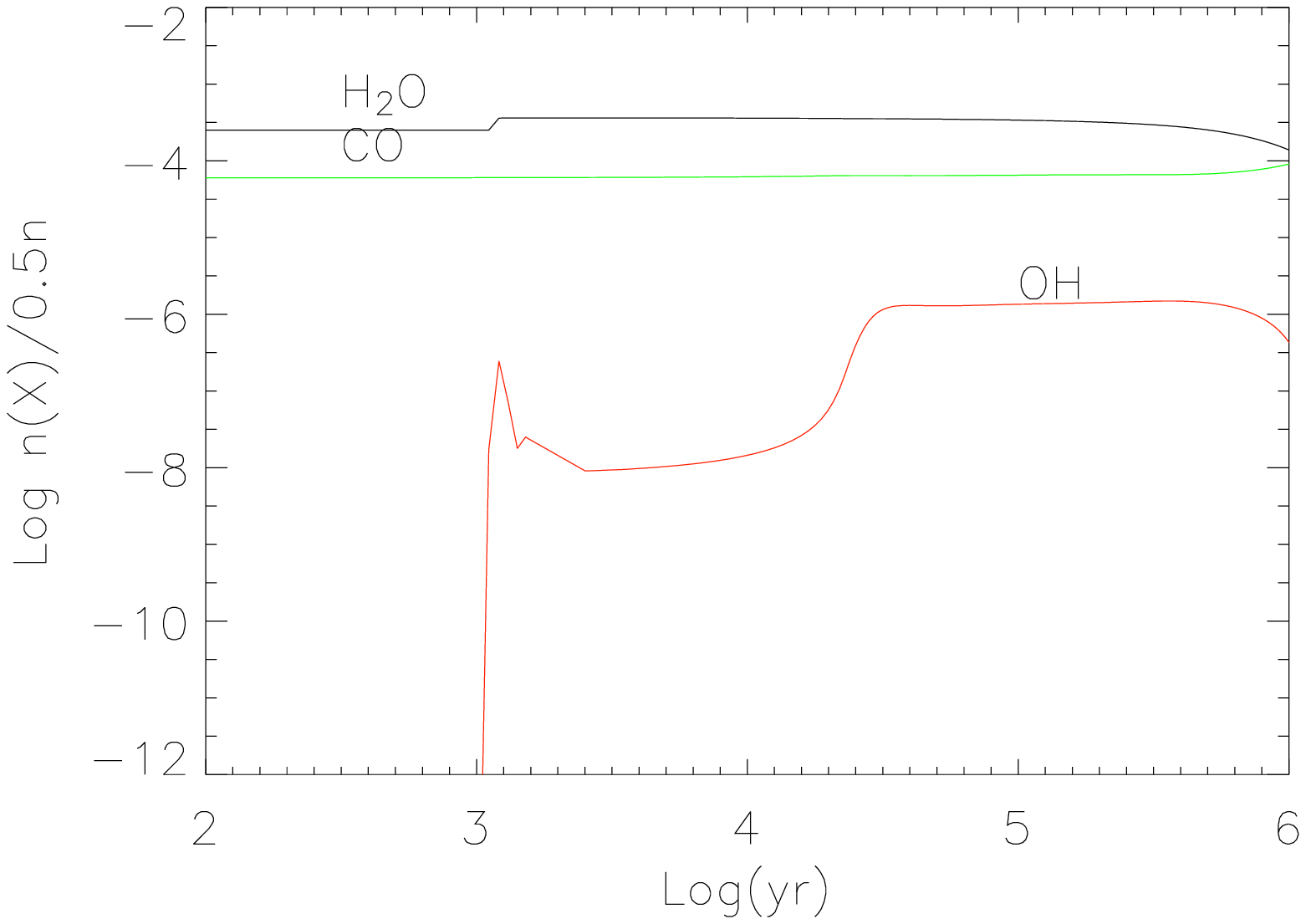}
\includegraphics[width=8cm,height=8cm]{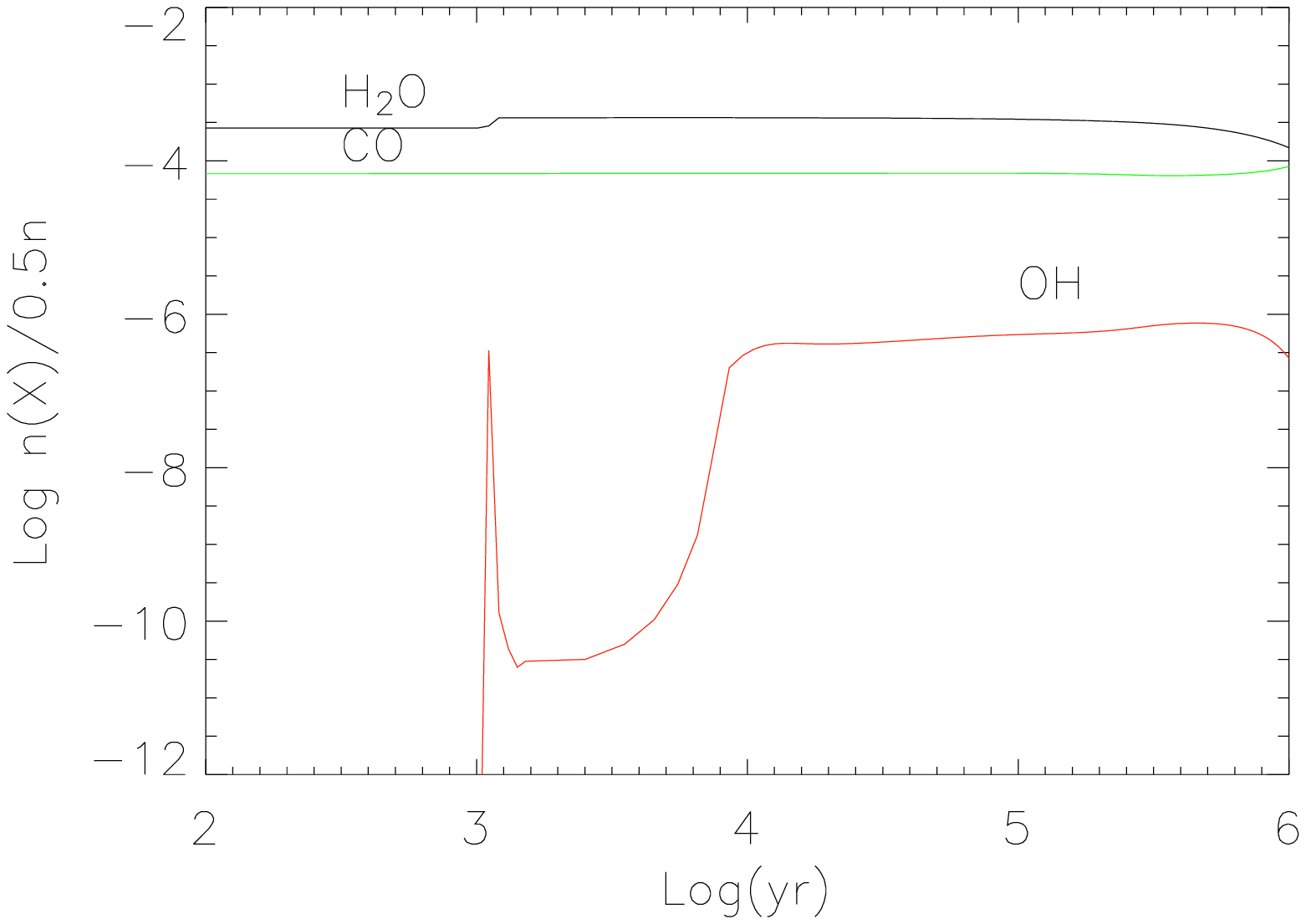}
    \caption{Plateau chemical models PL3 and PL14 (see Table~\ref{plat_para}) found to provide the best fit
    to the
    CO lines (see  also Figure~\ref{co_results}).}
        \label{PL1}
   \end{figure}

 \begin{figure}
    \centering
\includegraphics[width=8cm,height=8cm]{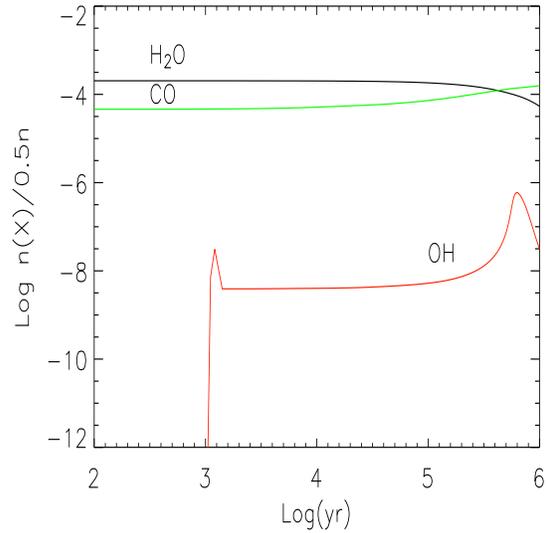}
    \caption{Model PL1, found to fit CO transitions with J$_{up}<$18 (see  also Figure~\ref{co_results}).}
        \label{PL2}
   \end{figure}

\subsection{The radiative transfer model} The chemical model
produces abundances that are used as inputs to the radiative
transfer model SMMOL (Rawlings \& Yates, 2001; van Zadelhoff et
al, 2002), along with values for the scale, density, dust temperature
and interstellar radiation field. The SMMOL code has accelerated
$\Lambda$-iteration (ALI) that solves the radiative transfer
problem in multi-level non-local conditions. It starts by calculating
the level population assuming LTE and the interstellar radiation
field as input continuum and then recalculates the total radiation
field and repeats the process until convergence is achieved. The
resulting emergent intensity distribution is then transformed to
the LWS units of flux (W cm$^{-2}$ $\mu$m$^{-1}$) taking into account the
different beam sizes (slightly different for each detector) and
convolved with an instrumental line profile corresponding to a 33
km s$^{-1}$ FWHM Lorentzian (Polehampton et al. 2007), in order to
directly compare with the
observations. The main input parameters include molecular data
such as molecular mass, energy levels, radiative and collisional
rates and also the dust size distribution and opacity. The CO
molecular data were taken from molecular public databases
(M\"{u}ller et al., 2001; Sch\"{o}ier et al., 2005). The input
parameters also include the gas and dust temperatures of the
object to model, such as the kinetic and dust thermal temperature. \\

\begin{figure}
    \centering
\includegraphics[width=8cm,height=8cm]{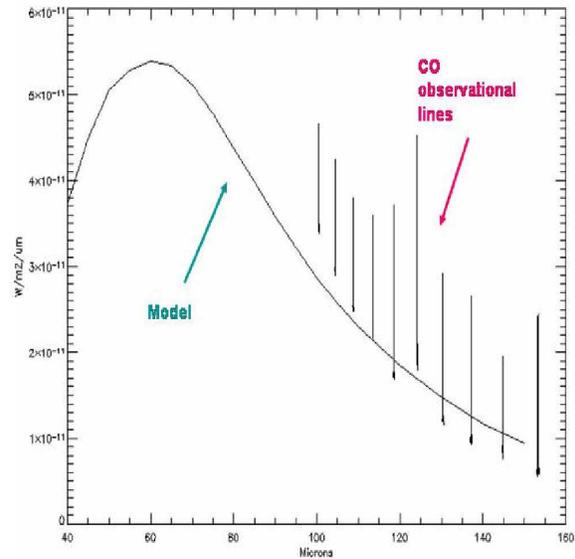}

 \caption{Model of the continuum adopted in the radiative transfer model.}
        \label{conti}

   \end{figure}

We estimated the dust temperature from the LWS grating observation
of Orion-KL (Lerate et al. 2006), fitting a black body function of
80 K. The gas kinetic temperature is based on the chemical model
input, which varies for the different models considered (Hot Core,
Plateau, PDR, see also Table~\ref{plat_para}). In order to
reproduce the observed continuum flux level, we adopted a
radiation field equivalent to 10$^{4}$ Habings. This value is in
agreement with previous continuum studies performed in the
submillimeter which constrains the ISRF enhancement to 10$^{3}$ to
10$^{4}$ times the standard IRSF (Jorgensen et al. 2006). The
microturbulent velocity was set to 5 km s$^{-1}$ and different
expansion velocities from 15--40 km s$^{-1}$ were also considered
for the shocked gas in the Plateau models. The resulting model
continuum is plotted in Figure~\ref{conti} along with the observed
CO lines. We estimated an error of less than 30\% for the fit to
the continuum, being the percentage deviation below the observed
continuum for wavelengths up to 120~$\mu$m and above the observed
continuum for the longest wavelengths.

\section{Results}
Figures~\ref{HC},~\ref{PL1} and ~\ref{PL2} show the chemical
models that best fit our observations. The selection of species to
be considered was based on the most abundant molecules detected in
our far-IR survey towards Orion-KL (i.e., CO, H$_{2}$O and OH). We
used a $\chi^{2}$ fit to compare model line profiles with
observations. The modelled lines were rebinned to the same
spectral sampling as the observations before the $\chi^{2}$
fitting. Figure~\ref{co_results} shows the best-fit models
compared to the observed CO lines. Table~\ref{abundances} shows
the computed densities, temperatures and fractional abundances
obtained for the different components. The contributions to the
line fluxes and profiles from the different models described below
were summed to give the total fluxes listed in
Table~\ref{abundances} and the profiles shown in
Figure~\ref{co_results}. Line peaks are also listed on column
four. Quoted error are based on goodness of fit estimation and do
not include calibration errors. The overall velocity error should
include the calibration error of $\pm$ 5 km s$^{-1}$ discussed in
section 2.
 The main results are listed on
Table~\ref{results} and can be summarized as follows:
\begin{itemize}

\item CO transitions with J$_{up}<$18 (i.e lines at wavelengths
$>$ 144.78 $\mu$m and upper energy levels E$_{up}$ lower than
$\approx$ 945 K) are best reproduced by a Plateau chemical model (PL2 in
Table~1) with a density of 3$\times10^{5}$ cm$^{-3}$, reaching a maximum
temperature of 300~K.
This could correspond to Plateau winds within the extended warm components. Ridge models with lower
temperature and densities fail to reproduce these transitions.

\item CO transitions with 18$<$J$_{up}<$25 (i.e. lines at
wavelengths between 104.45 $\mu$m and 144.78 $\mu$m and with upper
level energies between $\approx$ 1795 K and 945 K) are best
reproduced by a Hot Core model with a mass of $\approx$ 25
M$\odot$, a density of 1$\times$ 10$^{7}$ cm$^{-3}$ and a diameter
of $\approx$ 0.020 pc.

\item Transitions with 30$>$J$_{up}>$25 (i.e lines between 87.19
and 104.45 $\mu$m, with upper level energies in the range
2567--1795 K) are best reproduced by a Plateau model (PL3) with a
density of 3$\times$10$^{5}$ cm$^{-3}$, simulating a shock that heats the gas to 500~K during a period of 100 yr at the age of 1000 yr.

\item Higher energy CO transitions, with J$_{up}>$30, are best fitted by
a Plateau model (PL14) with a higher density, 1$\times$10$^{6}$ cm$^{-3}$
and a freeze-out efficiency of 80\%,  simulating a post-shock region at
1000 K which heats the gas, at an age of 1000 yr. All Plateau
models had a diameter of $\approx$ 0.061 pc.
\end{itemize}

X(CO), the CO/H$_2$ fractional abundance, varies slightly as a function of
cloud depth, from 7.0 $\times$ 10$^{-5}$ to 4.7 $\times$
10$^{-5}$, with an enhancement in the hot core component. This is
an effect of the different densities, and therefore freeze-out
efficiencies.
\begin{table*}
\centering
\begin{tabular}{c c c c c c c c c c c}
\hline
J  & Flux & Predicted Flux  & Line peak& FWHM &Comp. Size & V$_{LSR}$ & [N(CO)/(H$_{2}$)]  & T$_{max}$  & Density\\
 &(10$^{-17}$W cm$^{-2}$) &(10$^{-17}$W cm$^{-2}$)&(km s$^{-1}$) & (km s$^{-1}$)&$\prime\prime$ &(km s$^{-1}$)& ($\times$ 10$^{-5}$)& (K)& (cm$^{-3}$)\\
 \hline\hline
16--15 & 9.48 $\pm$ 0.26 & 9.62 &  12.7 $\pm$ 0.4& 89.9 $\pm$ 11.5  & 50 & 9& 4.66 &300 &1$\times$10$^{5}$  \\
17--16 & 8.66 $\pm$ 0.13 & 8.78 &  14.6 $\pm$ 0.2&  75.2 $\pm$ 12.5&50 &9& 4.66 &300 &1$\times$10$^{5}$  \\
18--17 & 3.98 $\pm$ 0.26 & 4.08 & 15.5 $\pm$ 1.1 &56.5$\pm$ 11.2   &50 &9& 4.66 &300 &1$\times$10$^{5}$ \\
19--18 & 5.22 $\pm$ 0.05 & 5.41 & 18.4 $\pm$ 0.2 &55.4 $\pm$ 11.1  &10& 3--5& 6.99 & 200&1$\times$10$^{7}$ \\
20--19 & 6.19 $\pm$ 0.22 & 6.08 &  16.1 $\pm$ 0.6&50.2 $\pm$ 10.9  &10&3--5& 6.99 &200 & 1$\times$10$^{7}$\\
21--20 & 7.89 $\pm$ 0.58 & 7.68 &  15.0 $\pm$ 1.1&50.5 $\pm$ 10.8  &10&3--5& 6.99 &200 & 1$\times$10$^{7}$\\
22--21 & 6.58 $\pm$ 0.09 & 6.49 & 12.5 $\pm$ 0.2 &52.5 $\pm$ 10.9  &10&3--5& 6.99 & 200& 1$\times$10$^{7}$\\
23--22 & 4.82 $\pm$ 0.15 & 4.65 & 10.4 $\pm$ 6.3 & 53.2 $\pm$ 11.2 &10&3--5& 6.99 & 200& 1$\times$10$^{7}$\\
24--23 & 3.28 $\pm$ 0.12 & 3.20 & 9.7 $\pm$ 0.4 &  52.5 $\pm$ 10.8  &10&3--5& 6.99 & 200& 1$\times$10$^{7}$\\
25--24 & 3.67 $\pm$ 0.10 & 3.57 & 8.32 $\pm$ 0.2 & 56.6 $\pm$ 11.3  &10&3--5& 6.99 & 200& 1$\times$10$^{7}$\\
26--25 & 3.44 $\pm$ 0.06 & 3.31 & 11.5 $\pm$ 0.2 &  70.2 $\pm$ 11.6 &10&3--5& 6.99 & 200& 1$\times$10$^{7}$\\
27--26 & 3.82 $\pm$ 0.11 & 3.55 & 9.2 $\pm$ 0.3&  70.1 $\pm$ 11.7   &30&7--8 &6.84   &1000  &1$\times$10$^{6}$ \\
28--27 & 1.89 $\pm$ 0.11 & 1.55 &  10.8 $\pm$ 0.6&  68.5 $\pm$ 11.5 &30&7--8&6.84 &1000  &1$\times$10$^{6}$ \\
29--28 & 1.58 $\pm$ 0.14 & 1.25 &  5.1 $\pm$ 0.4&  90.2$\pm$ 11.8  &30&7--8&6.84 &1000  &1$\times$10$^{6}$ \\
30--29 & 1.58 $\pm$ 0.11 & 1.15 & 2.8 $\pm$ 0.2 &  57.5 $\pm$ 10.5 &30&7--8&6.84 &  1000 & 1$\times$10$^{6}$\\
32--31 &  1.52 $\pm$ 0.09 & 0.85 & 10.8 $\pm$ 0.6&  79.5 $\pm$ 11.6 &30& 7--8&6.03 &500 & 3$\times$10$^{5}$\\
33--32 & 2.32 $\pm$ 0.20 & 0.81 & 3.7 $\pm$ 0.3&   99.8$\pm$12.1  &30& 7--8&6.03 &500 & 3$\times$10$^{5}$\\
34--33 & 0.60 $\pm$ 0.09 & 0.58 & 4.9 $\pm$ 0.7&   66.6 $\pm$ 10.2  &30& 7--8&6.03 &500 & 3$\times$10$^{5}$\\
35--34 & 3.46 $\pm$ 0.40 & 0.78 & 0.1 $\pm$ 0.01 &  120.2$\pm$15.5 &30&7--8&6.03 &500 & 3$\times$10$^{5}$\\
36--35 & 0.62 $\pm$ 0.13 & 0.51 & 1.6 $\pm$ 0.3 &   80.2$\pm$ 11.3 &30&7--8&6.03 &500 & 3$\times$10$^{5}$\\
37--36 & 0.70 $\pm$ 0.06 & 0.59 & 10.1 $\pm$ 0.9 &  60.5 $\pm$ 11.5 &30&7--8&6.03 &500 & 3$\times$10$^{5}$\\
38--37 & 0.45 $\pm$ 0.14 & 0.29 &  15.5 $\pm$ 4.8&   56.6 $\pm$ 11.6&30&7--8&6.03 &500 & 3$\times$10$^{5}$\\
39--38 & 0.48 $\pm$ 0.14 & 0.31 & 12.4 $\pm$ 3.6 &   54.2 $\pm$ 11.9&30&7--8&6.03 &500 & 3$\times$10$^{5}$\\

\hline

\end{tabular}
\caption{CO line fluxes and summary of the main model parameters
and results.} \label{abundances}
\end{table*}

\begin{table*}
\begin{tabular}{c c c c c c c}
\hline\hline Component & Size ($^{\prime\prime}$) & Density
(cm$^{-3}$) & T$_{gas}$ (K)& T$_{shock}$ (K)& mco ($\%$)& CO J
range \\
\hline
Extended warm gas & 30 & 3$\times$10$^{5}$ & 90 & 300 & 60 & $<$18\\
Hot Core & 10 & 1 $\times$ 10$^{7}$ & 200 & & 98 & 18--25\\
Plateau & 30 & 3$\times$10$^{5}$ & 90 & 500 & 60 & 25--30\\
Plateau & 30 & 1$\times$10$^{6}$& 100 & 1000 &80 & $>$30\\
\hline
\end{tabular}
 \caption{Summary of the main components and their physical
parameters of size, density,
 gas temperature, dust temperature and percentage of mantle CO from the CO line modelling.}
 \label{results}
\end{table*}

\begin{figure}
    \centering
   \includegraphics[width=4cm,height=4cm]{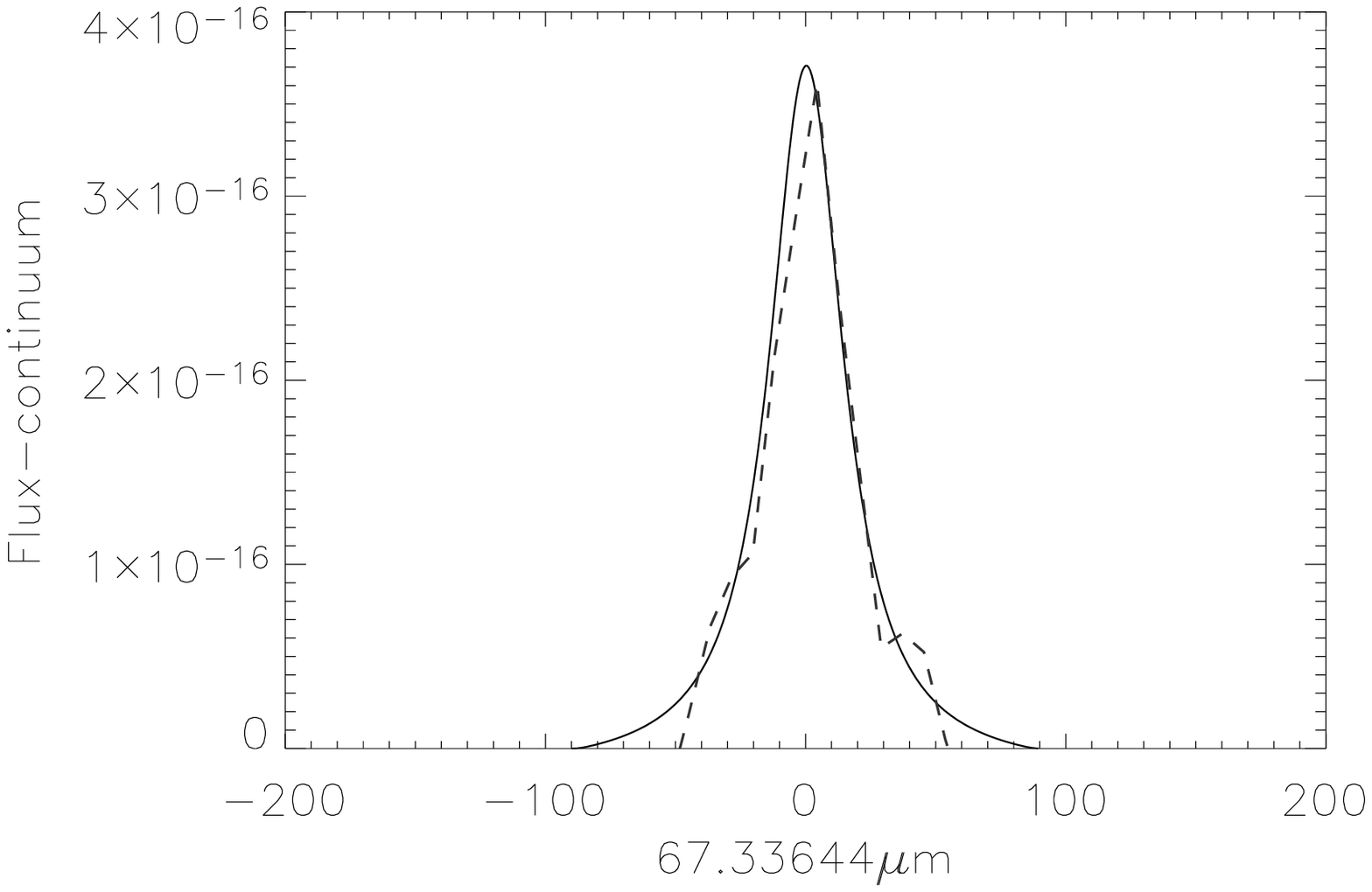}
\includegraphics[width=4cm,height=4cm]{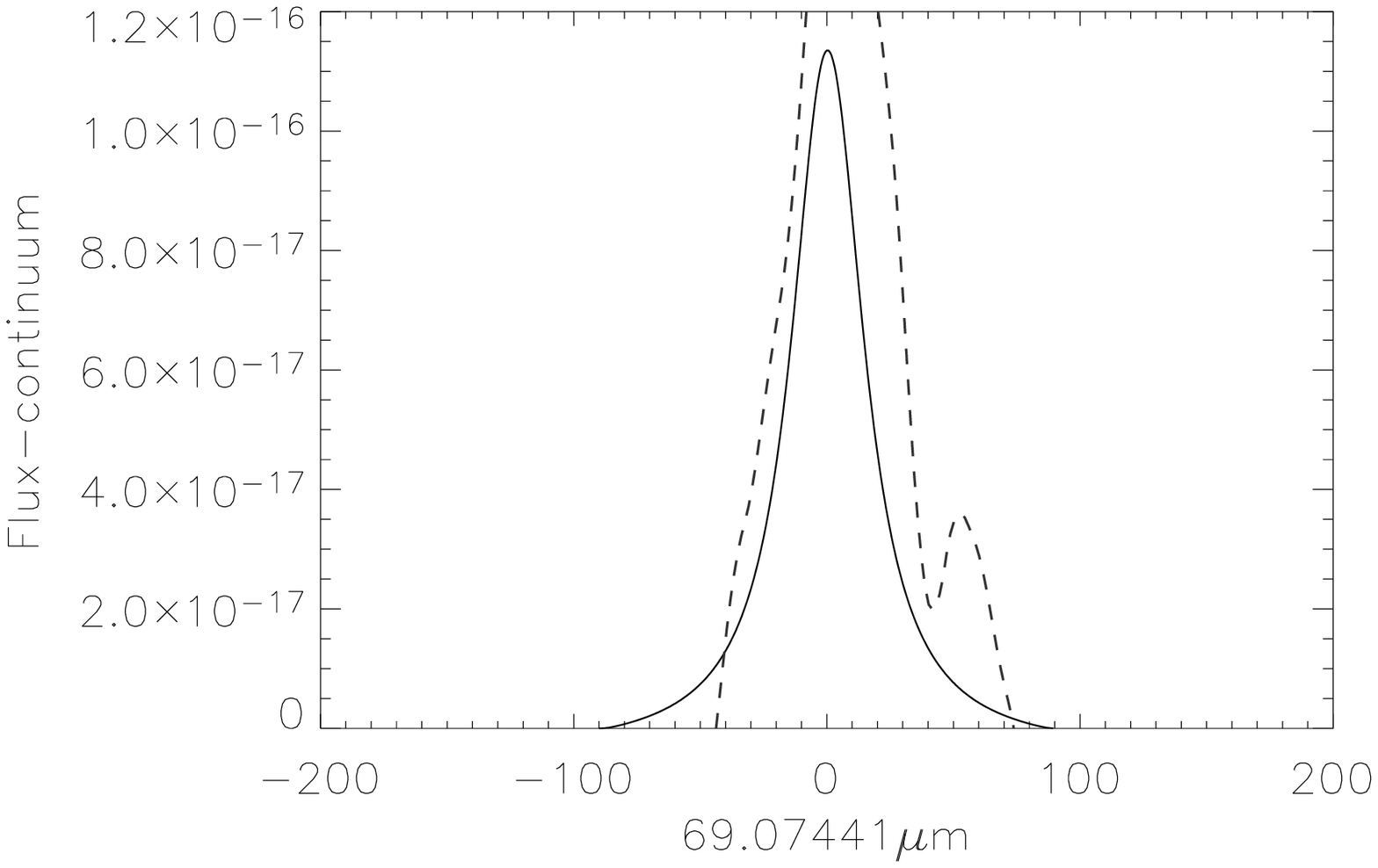}
\includegraphics[width=4cm,height=4cm]{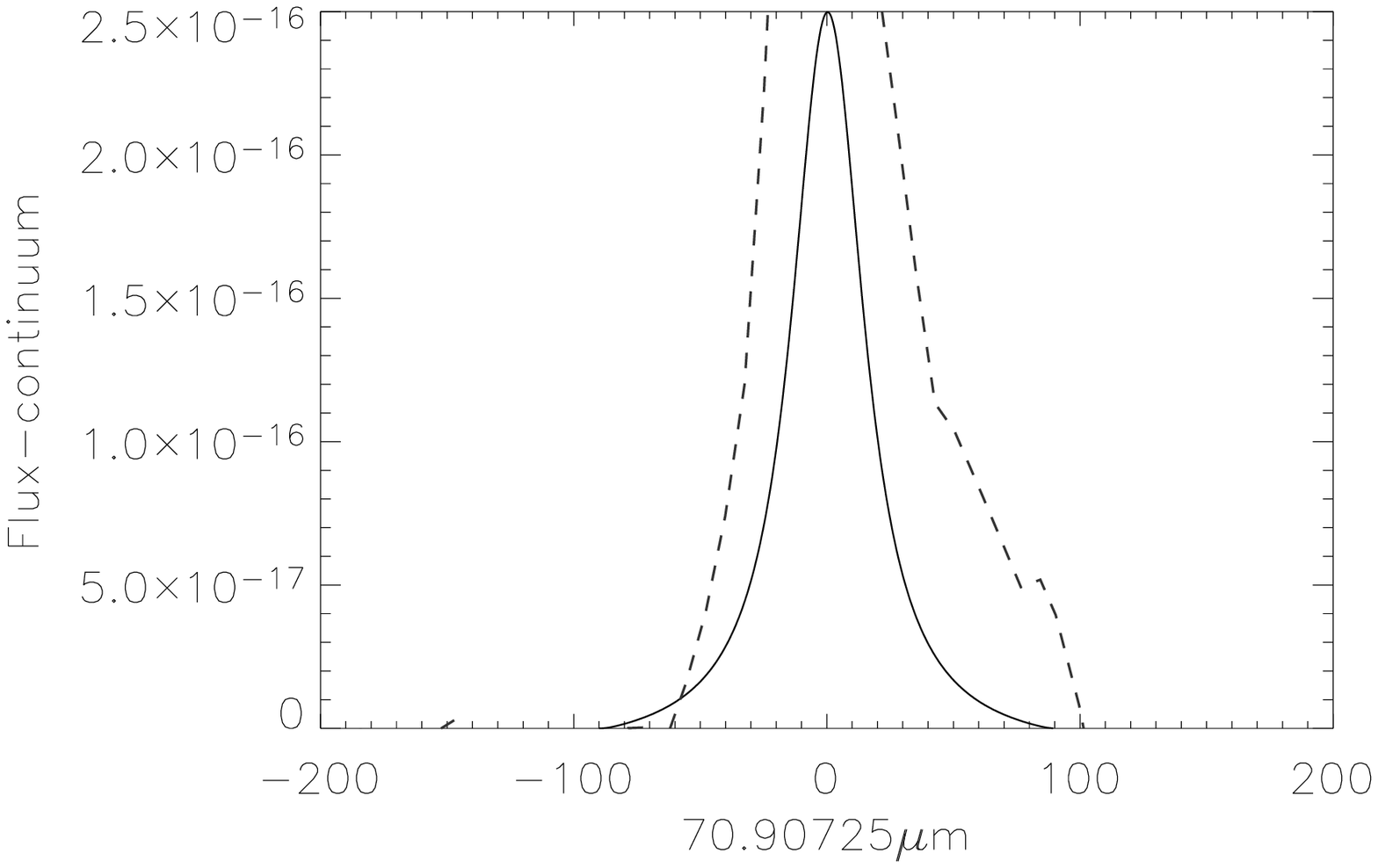}
\includegraphics[width=4cm,height=4cm]{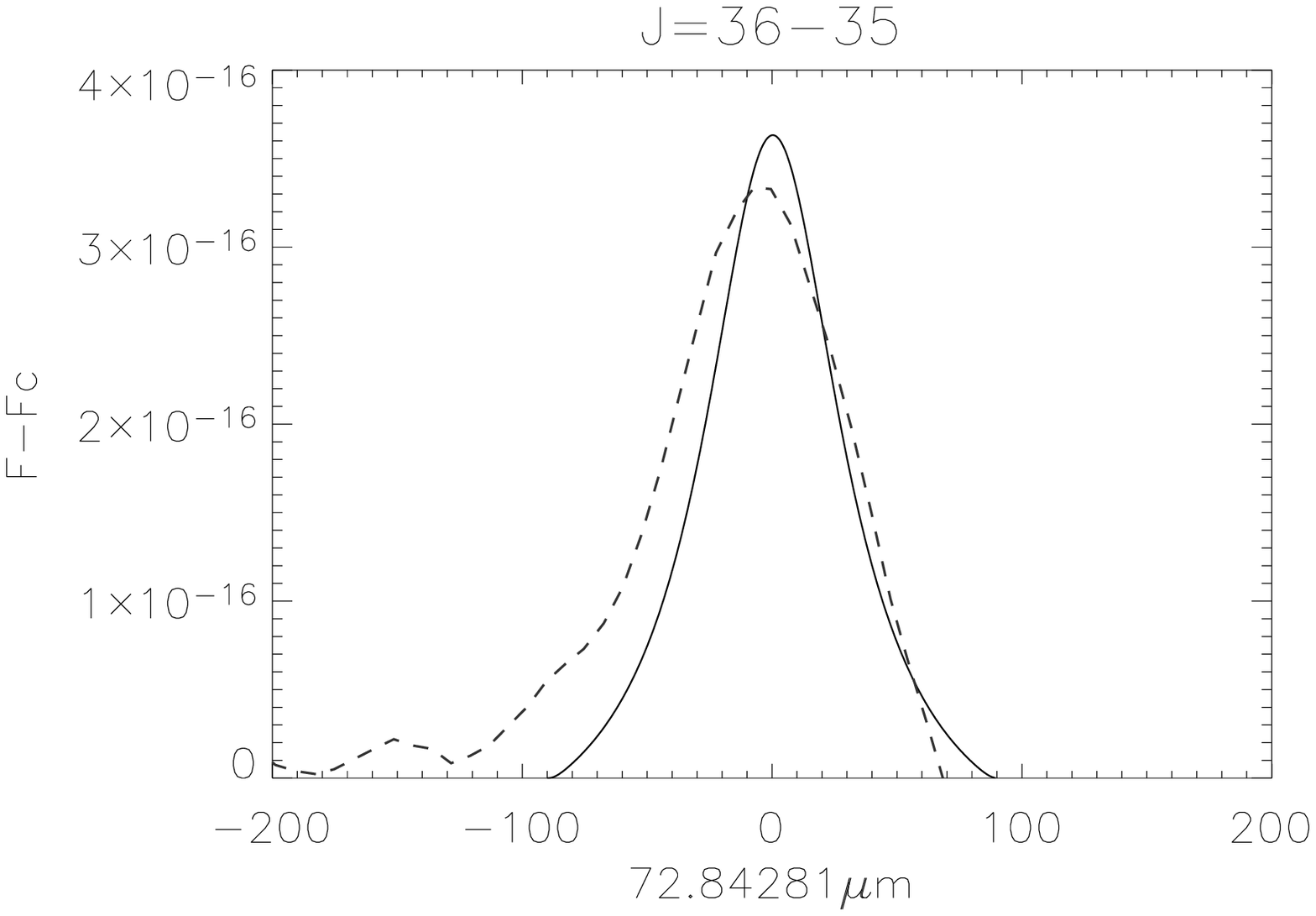}
\includegraphics[width=4cm,height=4cm]{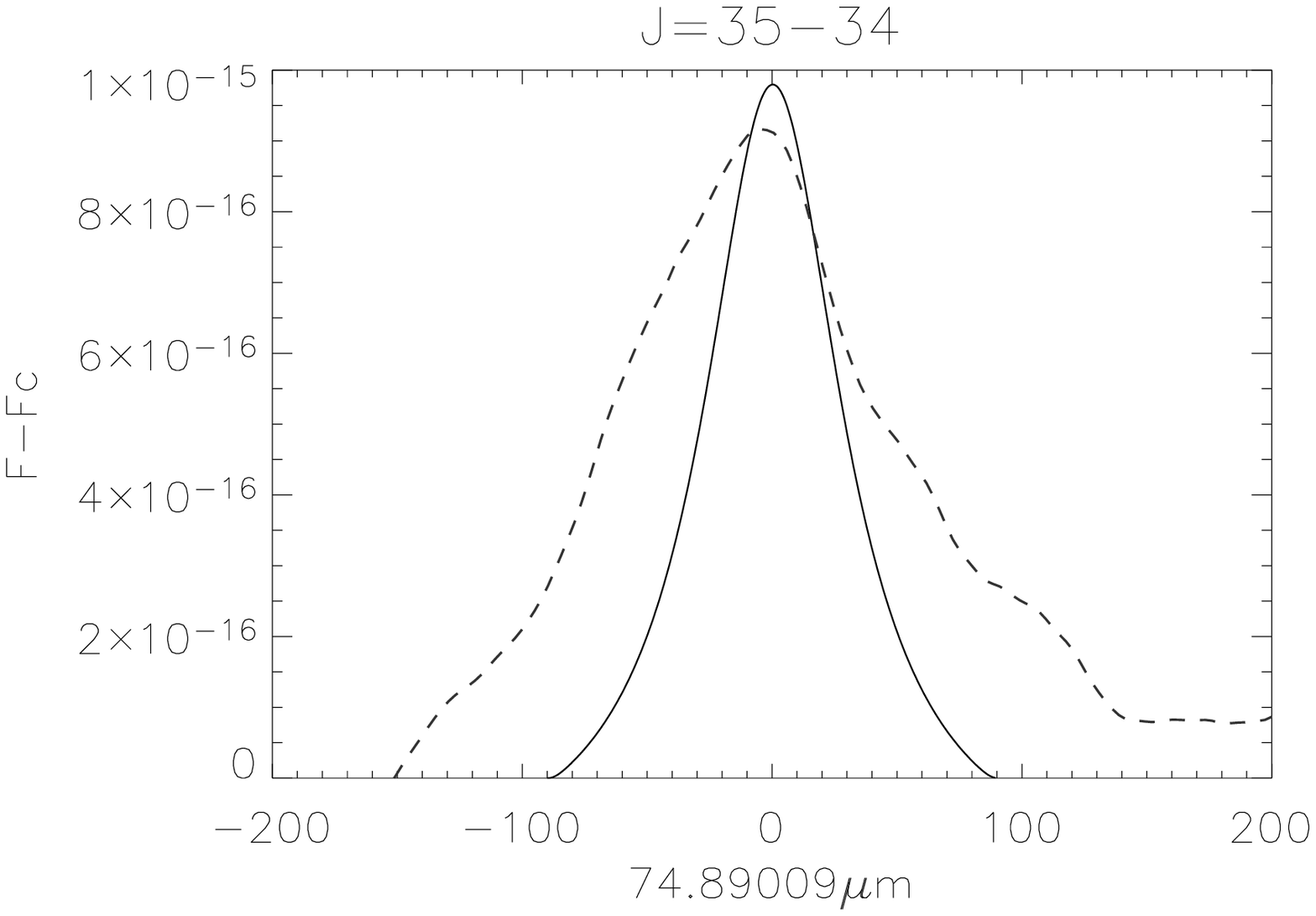}
\includegraphics[width=4cm,height=4cm]{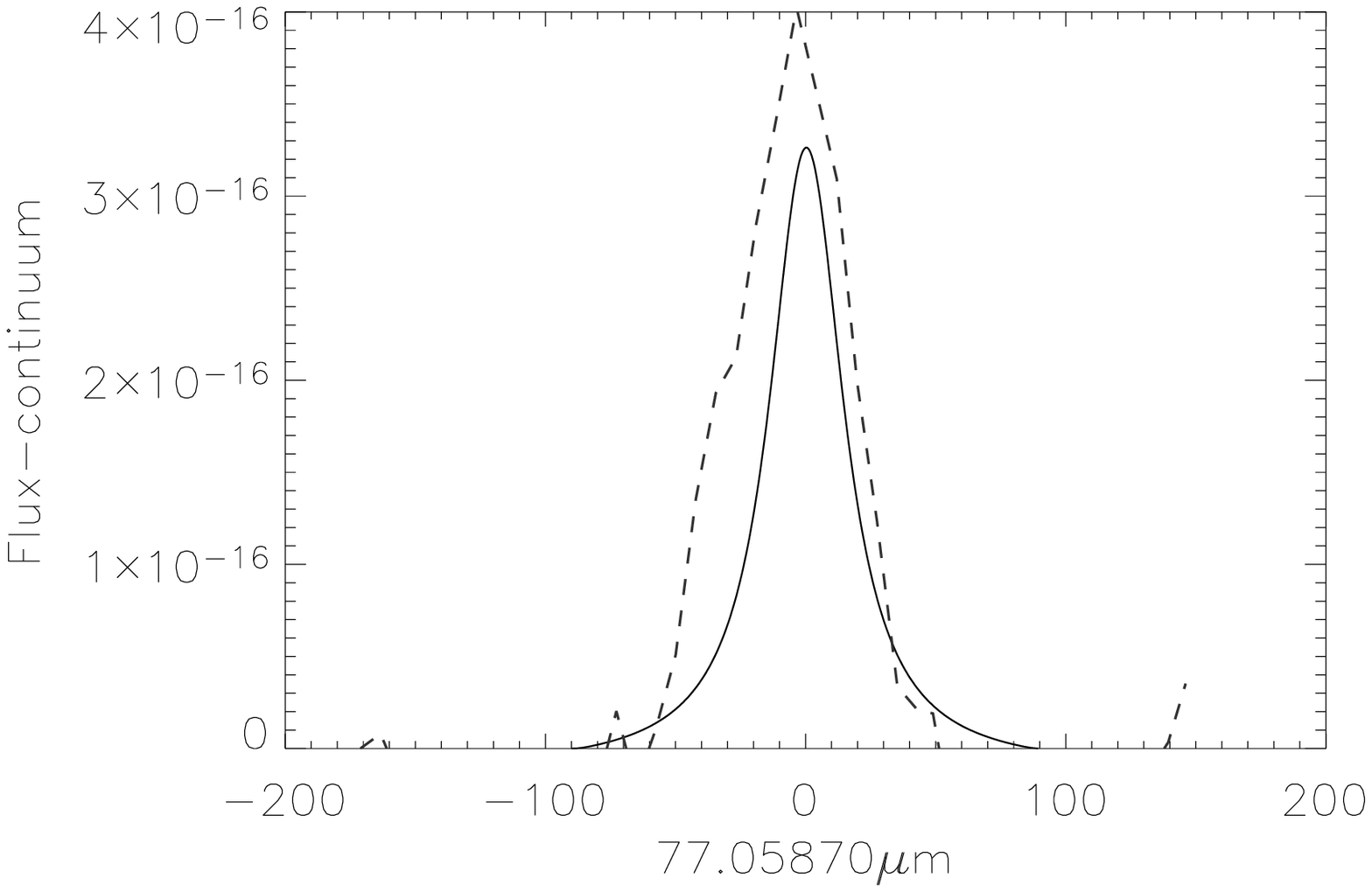}

\label{co_results} \caption{Line profiles from the radiative
transfer models (solid lines) over-plotted on the observed CO
lines (dotted lines). The ordinate corresponds to the
continuum-subtracted flux and the abscissa is the velocity in km
s$^{-1}$.
    CO transitions with 18$<$J$_{up}<$25
(i.e from wavelengths 104.45 to 144.78 $\mu$m) are reproduced by
the Hot Core model. Transitions with 30$>$J$_{up}>$25 (i.e lines
between 87.19 and 104.45 $\mu$m) are reproduced by the Plateau
model PL3. CO transitions with J$_{up}>$32 are reproduced by the
Plateau model PL14 and transitions with J$_{up}<$18 by model PL1
(see text for a description of these models)}
   \end{figure}
\addtocounter{figure}{-1}
\begin{figure}
    \centering

\includegraphics[width=4cm,height=4cm]{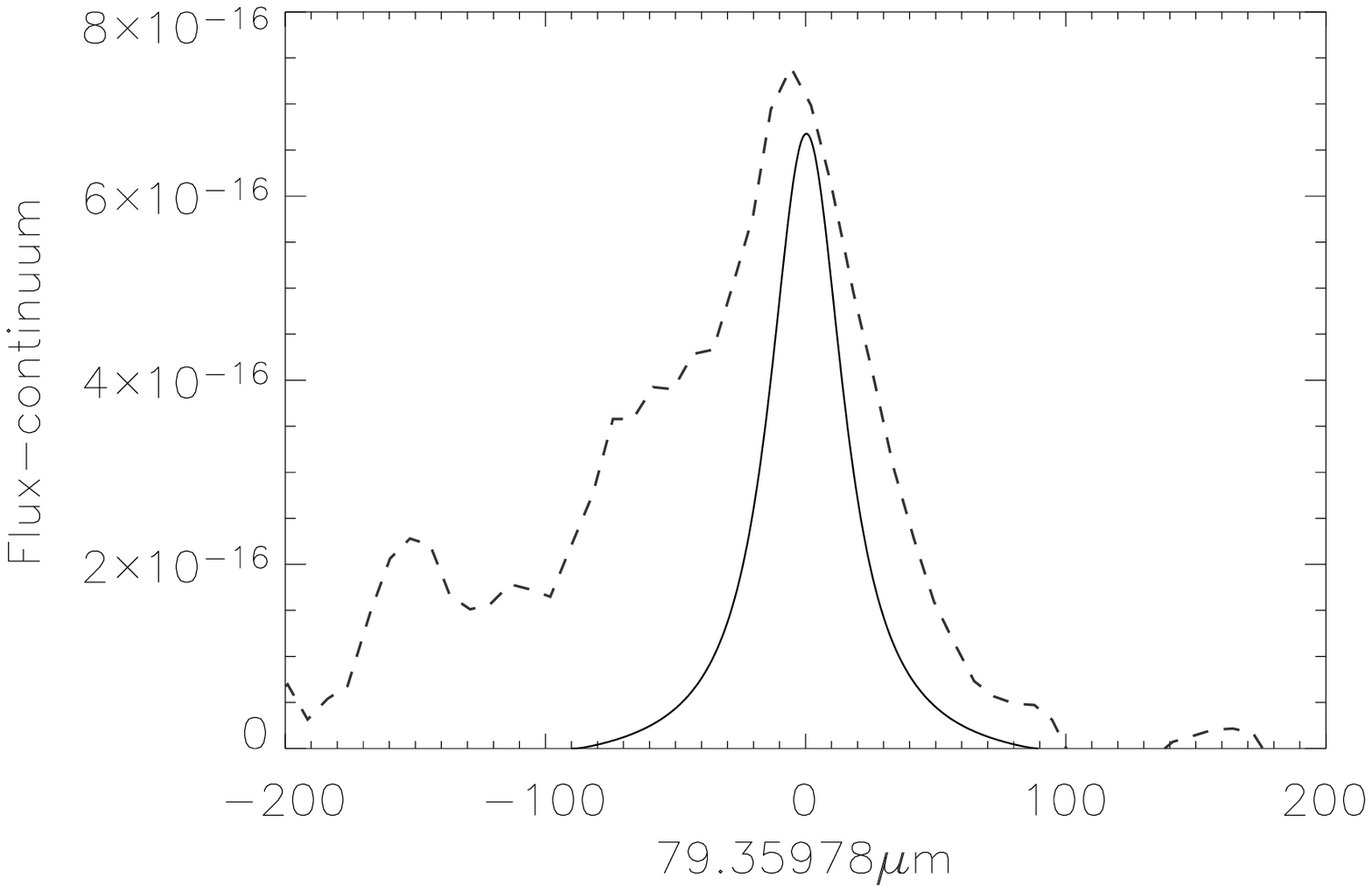}
\includegraphics[width=4cm,height=4cm]{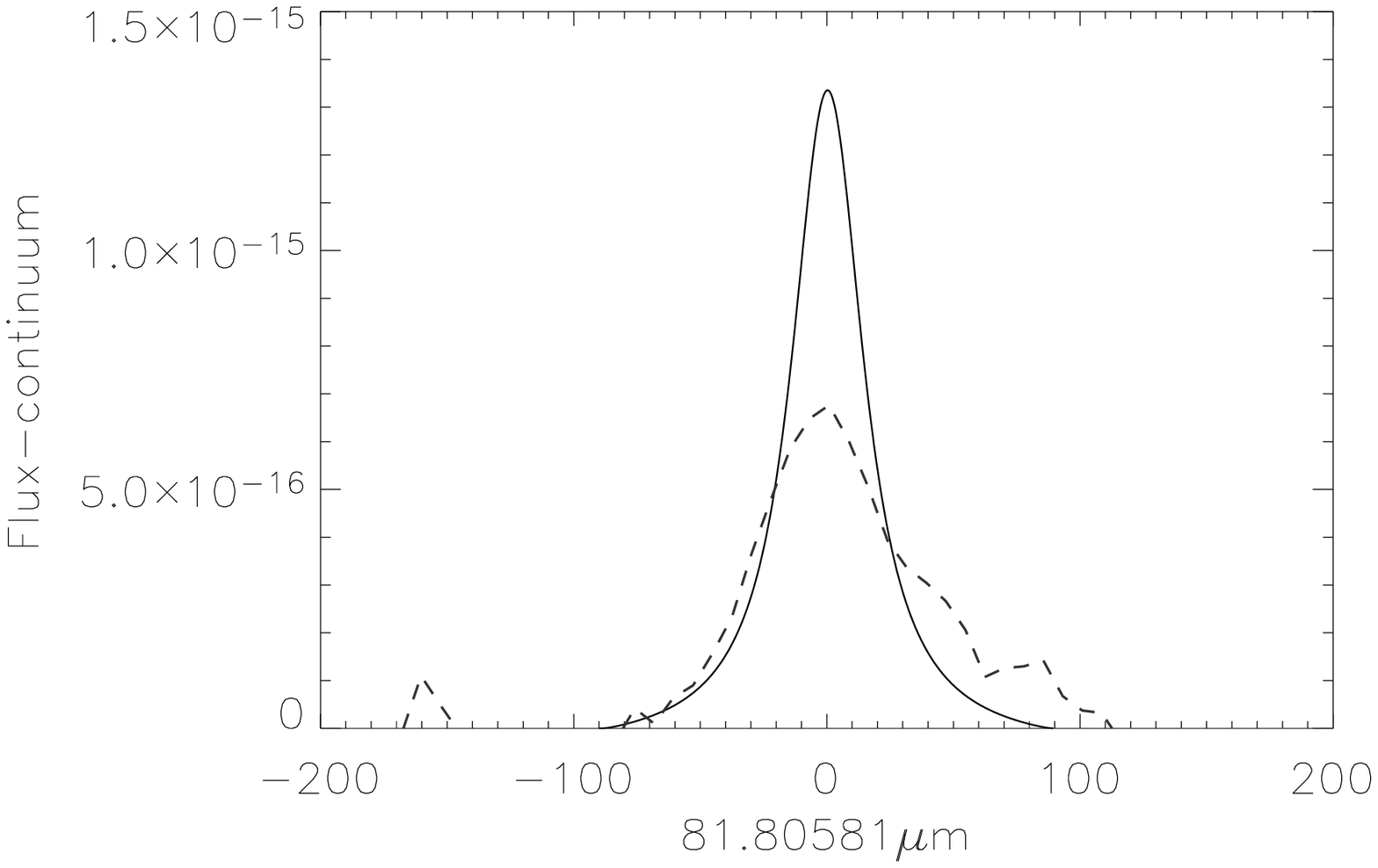}
\includegraphics[width=4cm,height=4cm]{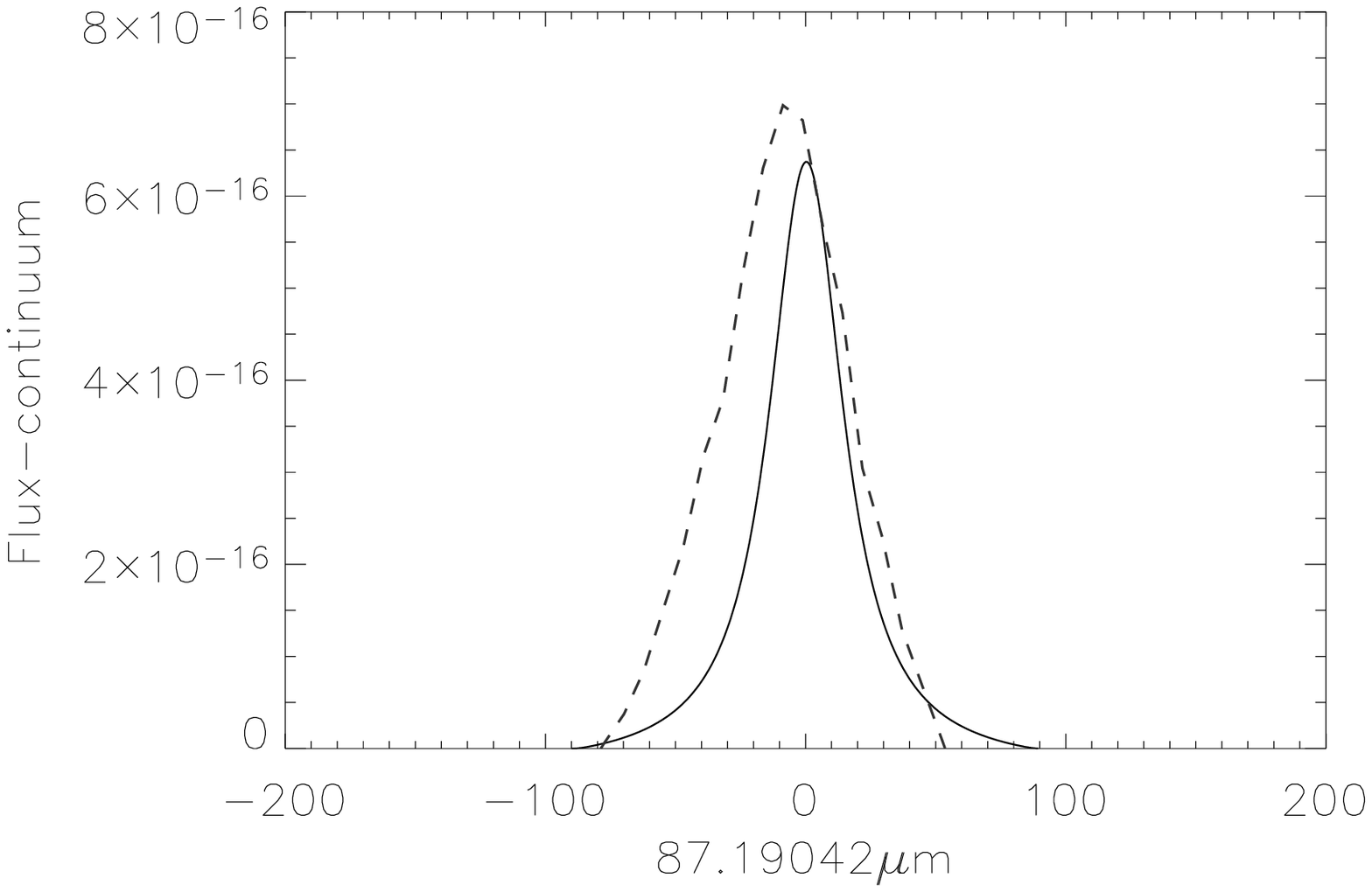}
\includegraphics[width=4cm,height=4cm]{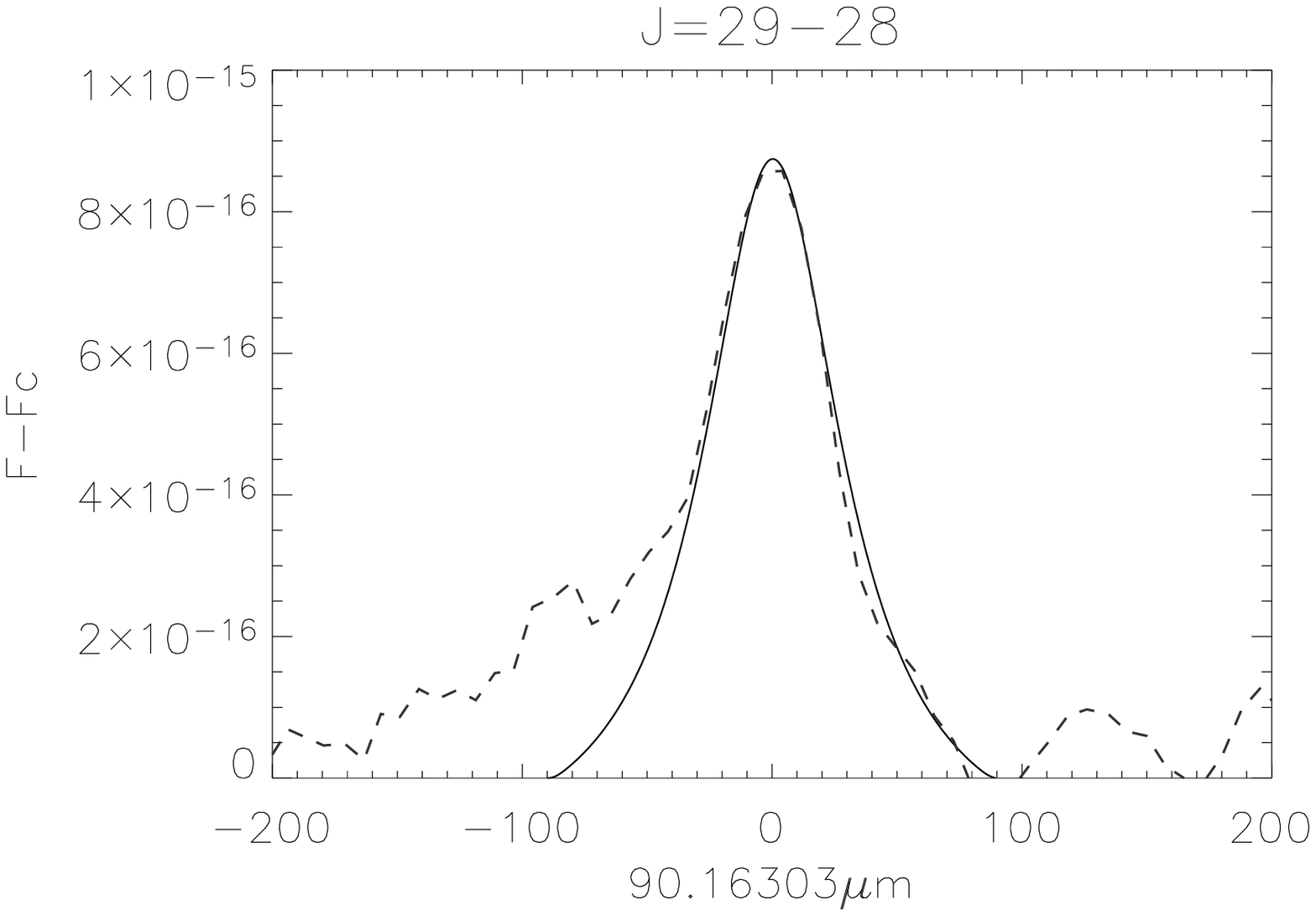}
\includegraphics[width=4cm,height=4cm]{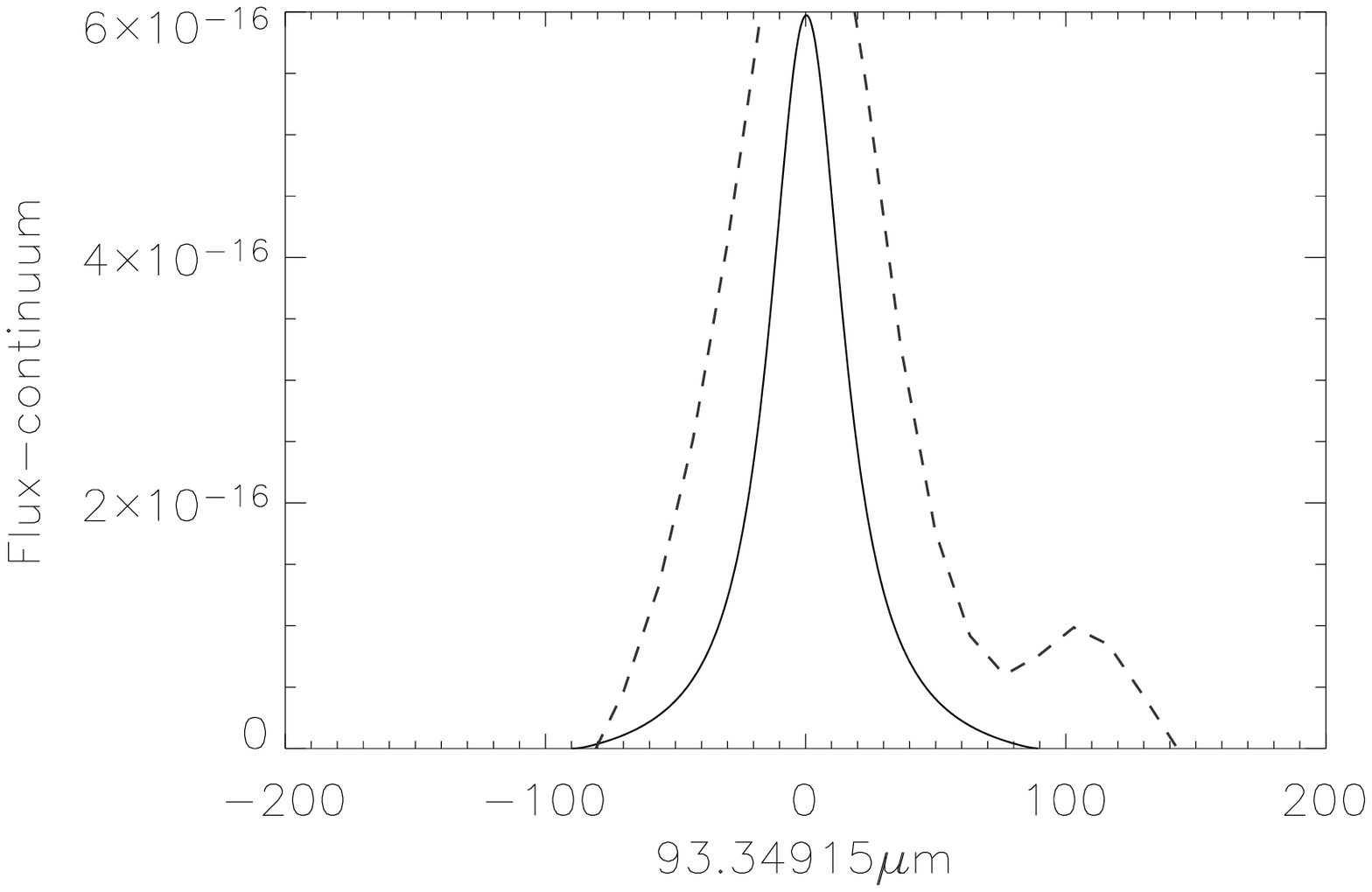}
\includegraphics[width=4cm,height=4cm]{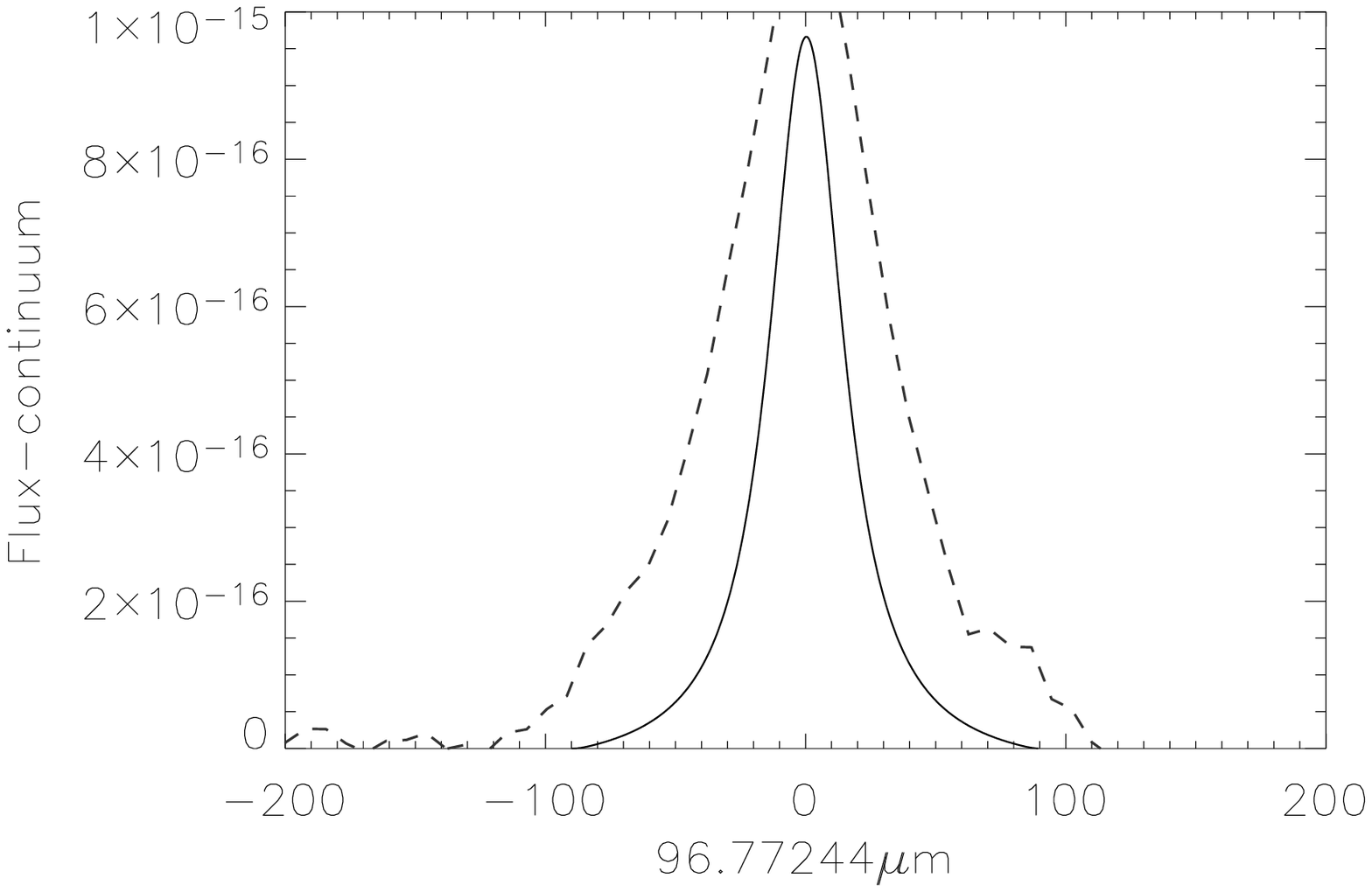}
\includegraphics[width=4cm,height=4cm]{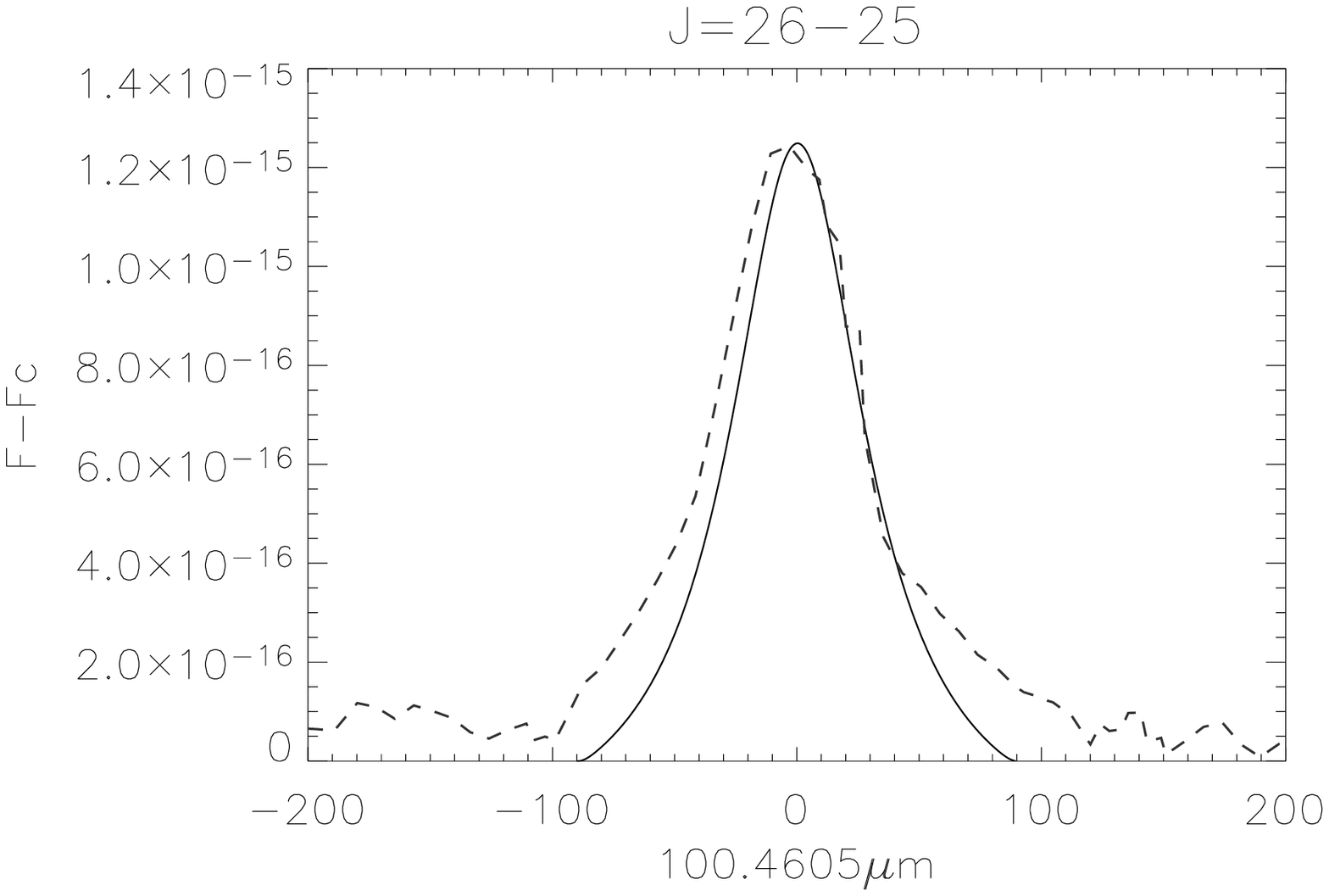}
\includegraphics[width=4cm,height=4cm]{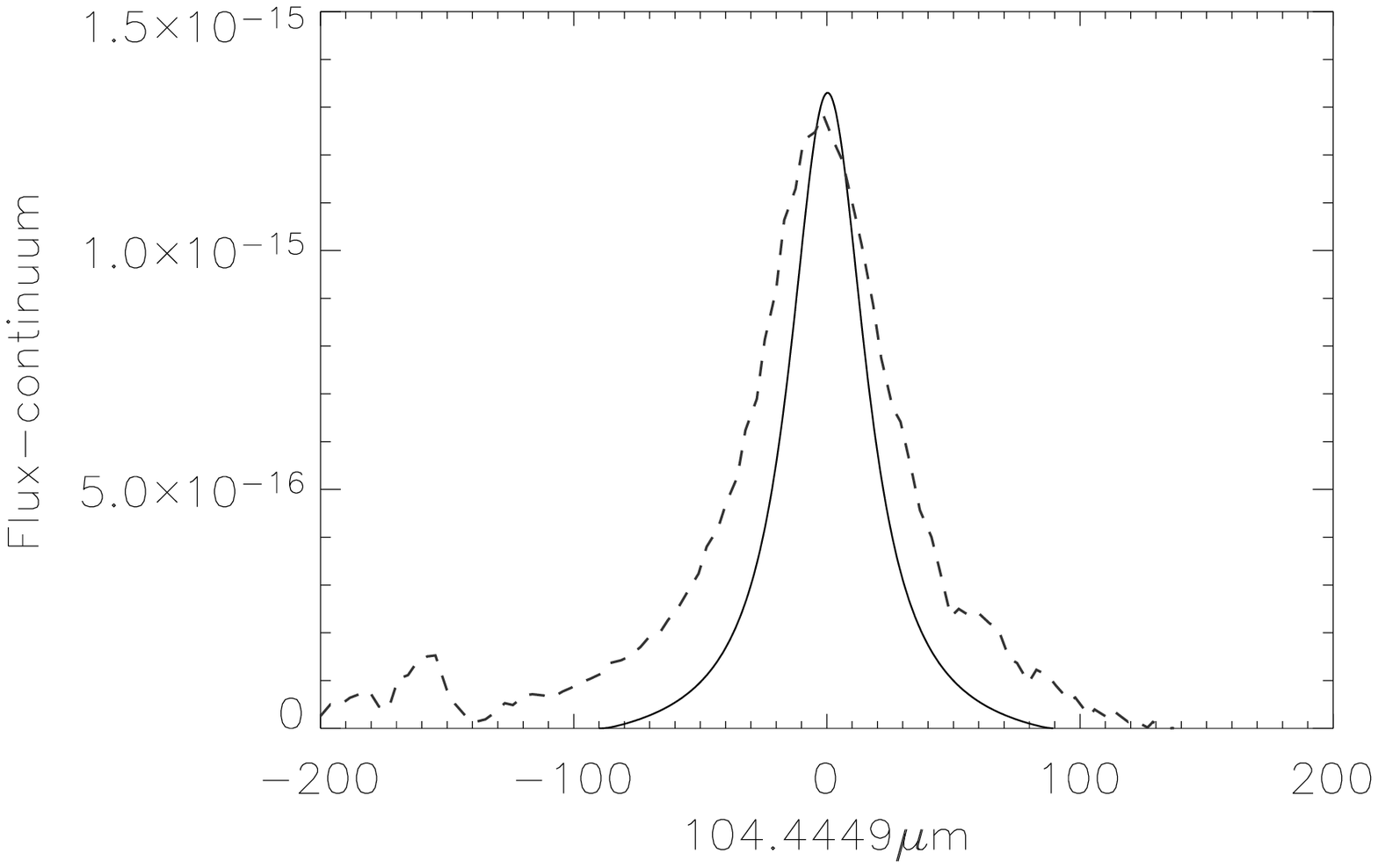}
 \caption{Continued}
 \end{figure}

\addtocounter{figure}{-1}
\begin{figure}
    \centering

\includegraphics[width=4cm,height=4cm]{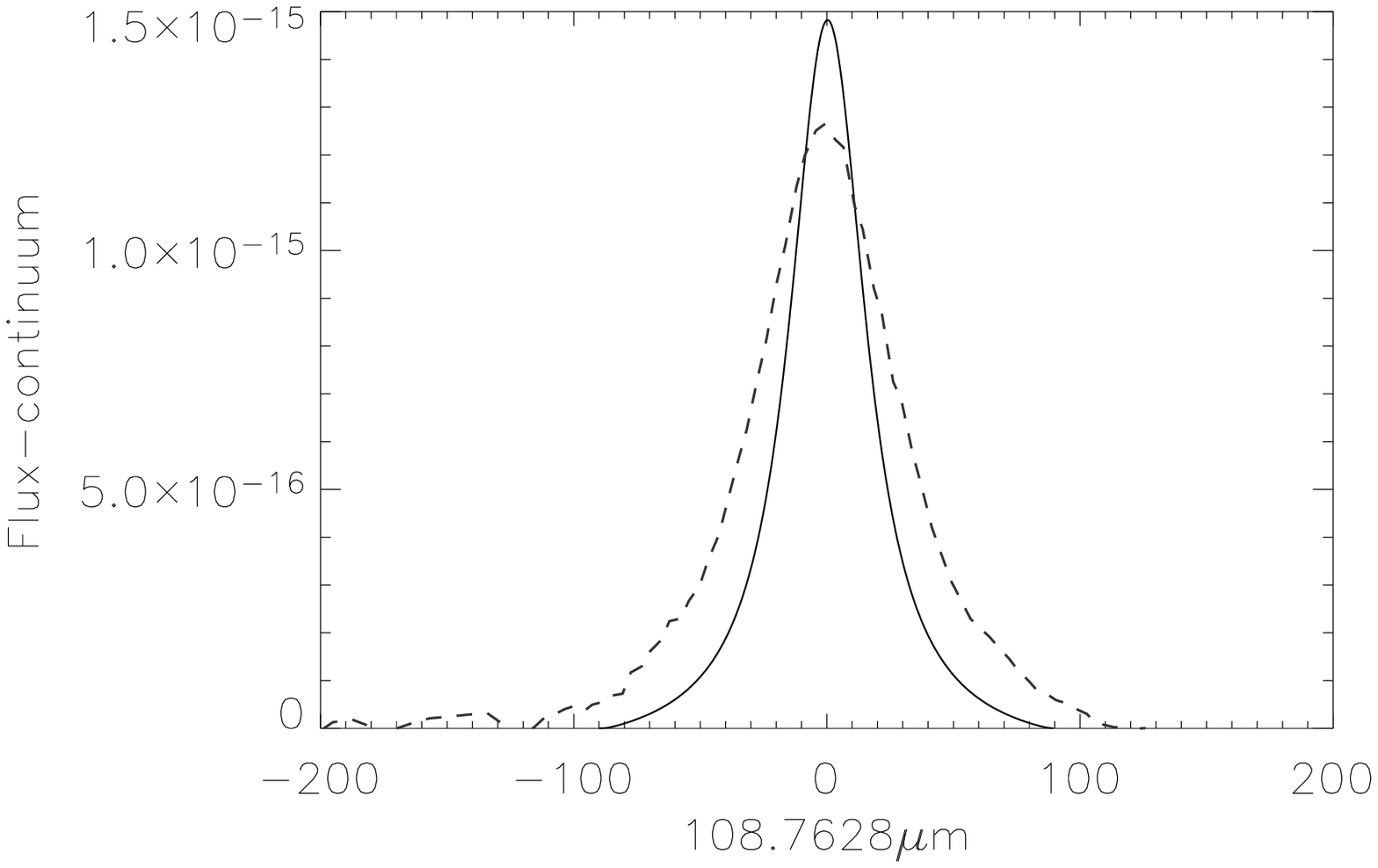}
\includegraphics[width=4cm,height=4cm]{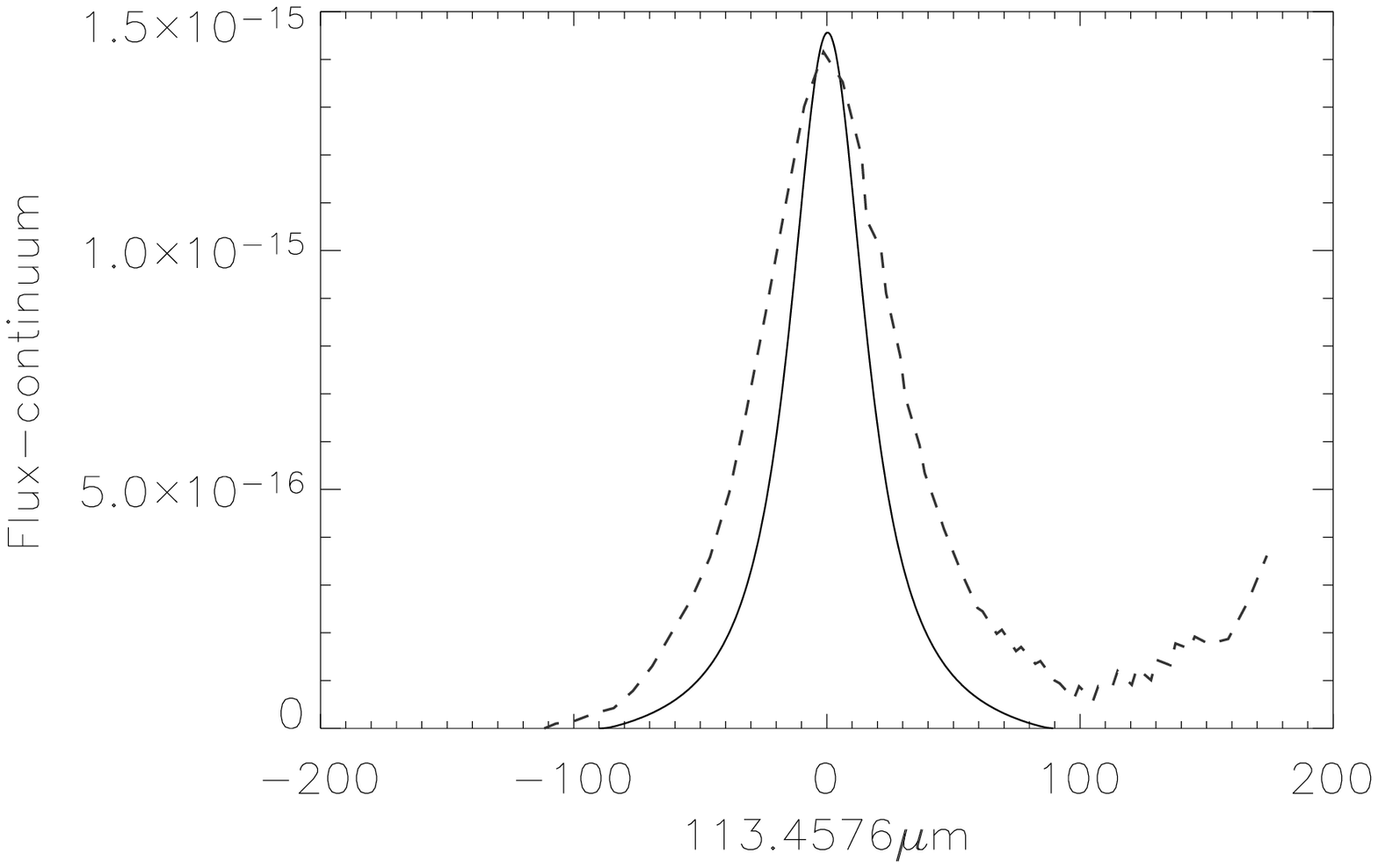}
\includegraphics[width=4cm,height=4cm]{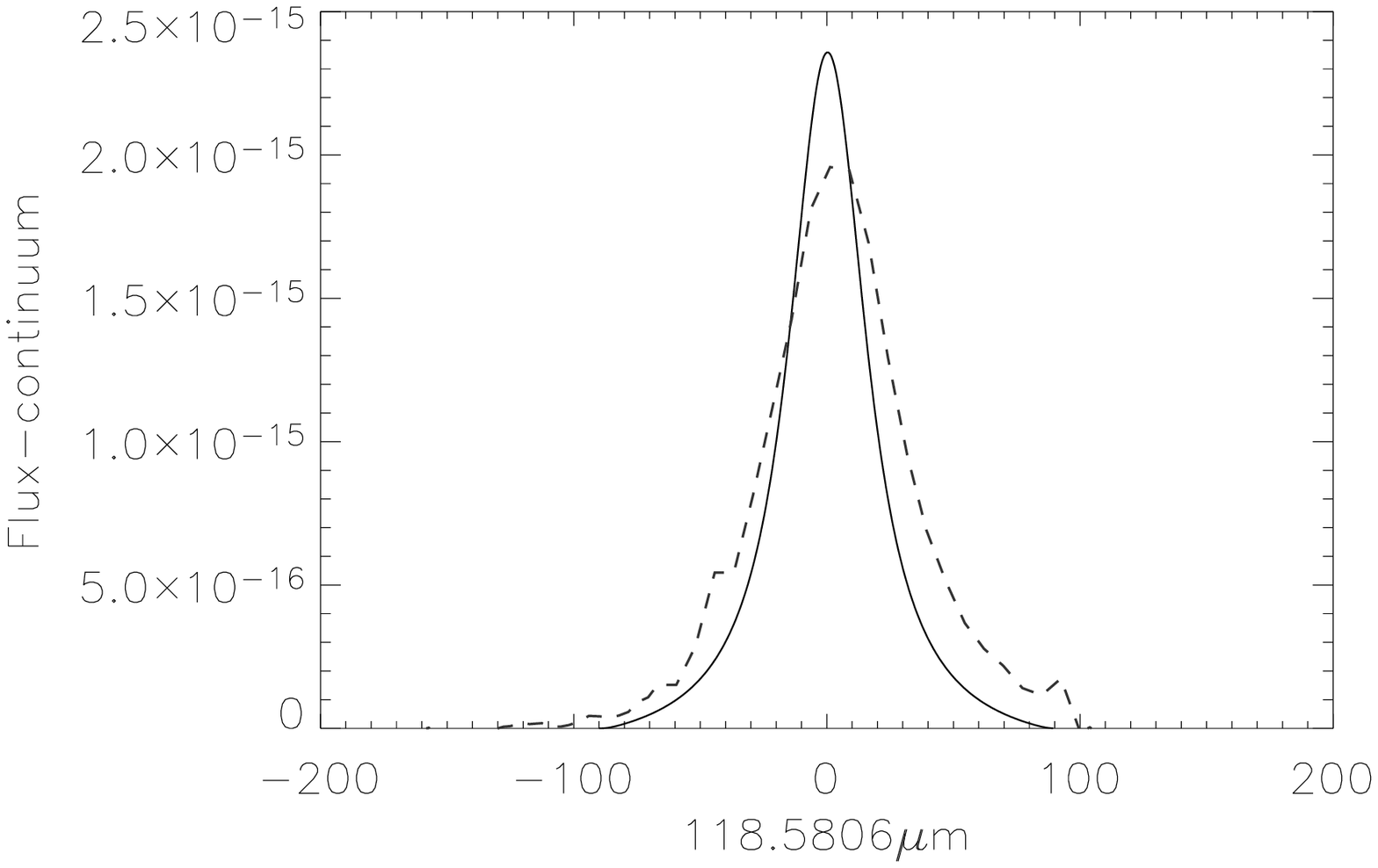}
\includegraphics[width=4cm,height=4cm]{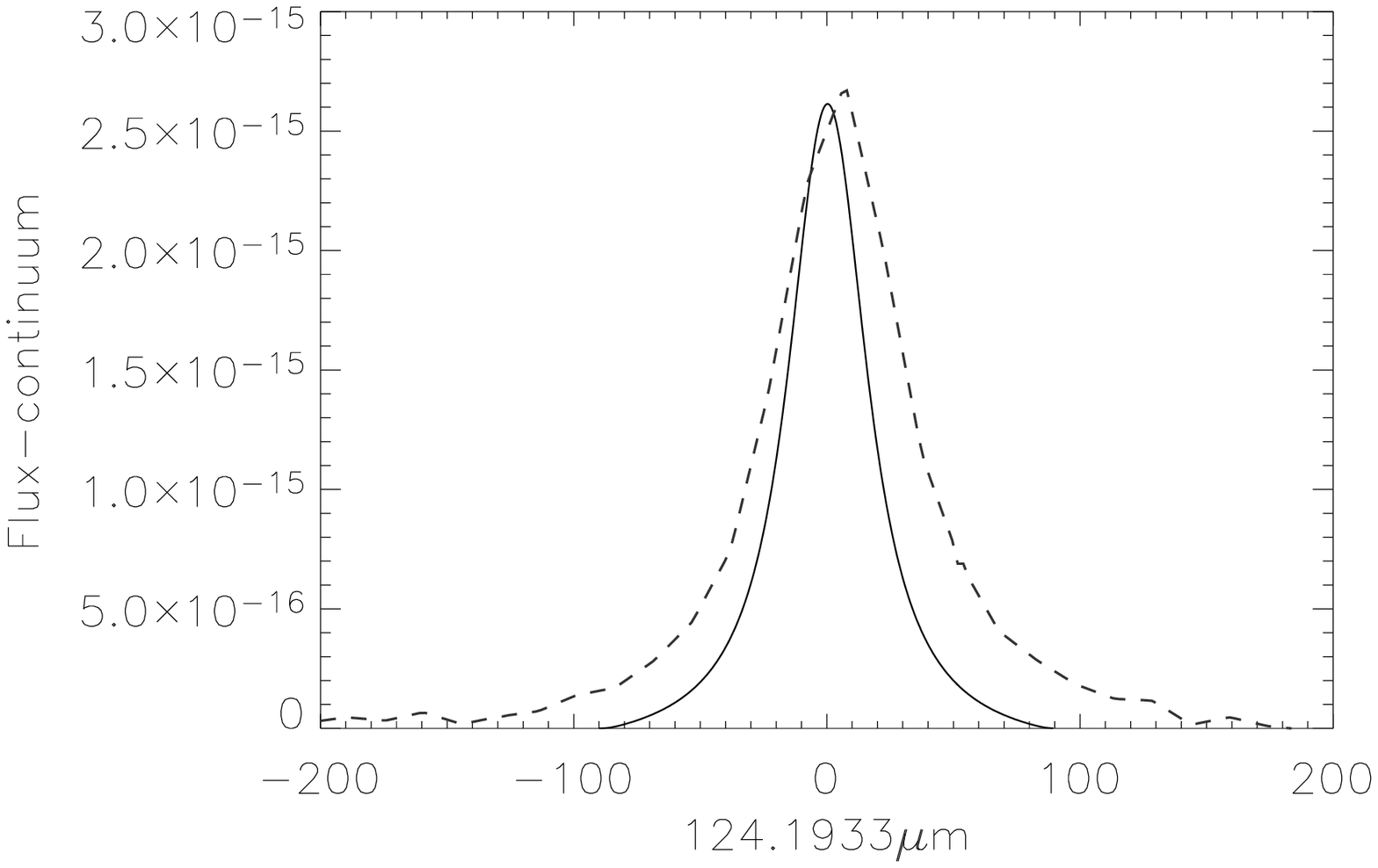}
\includegraphics[width=4cm,height=4cm]{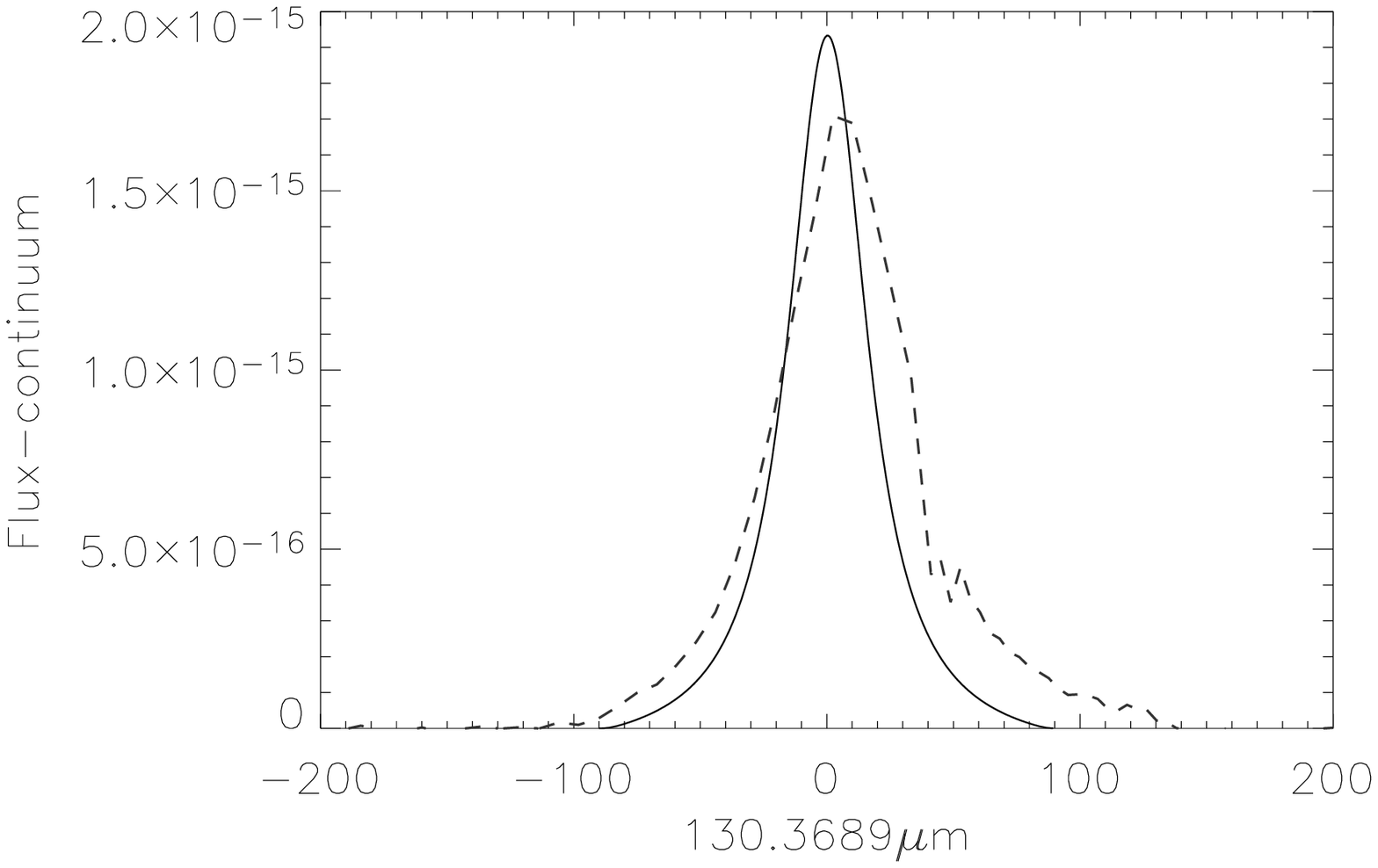}
\includegraphics[width=4cm,height=4cm]{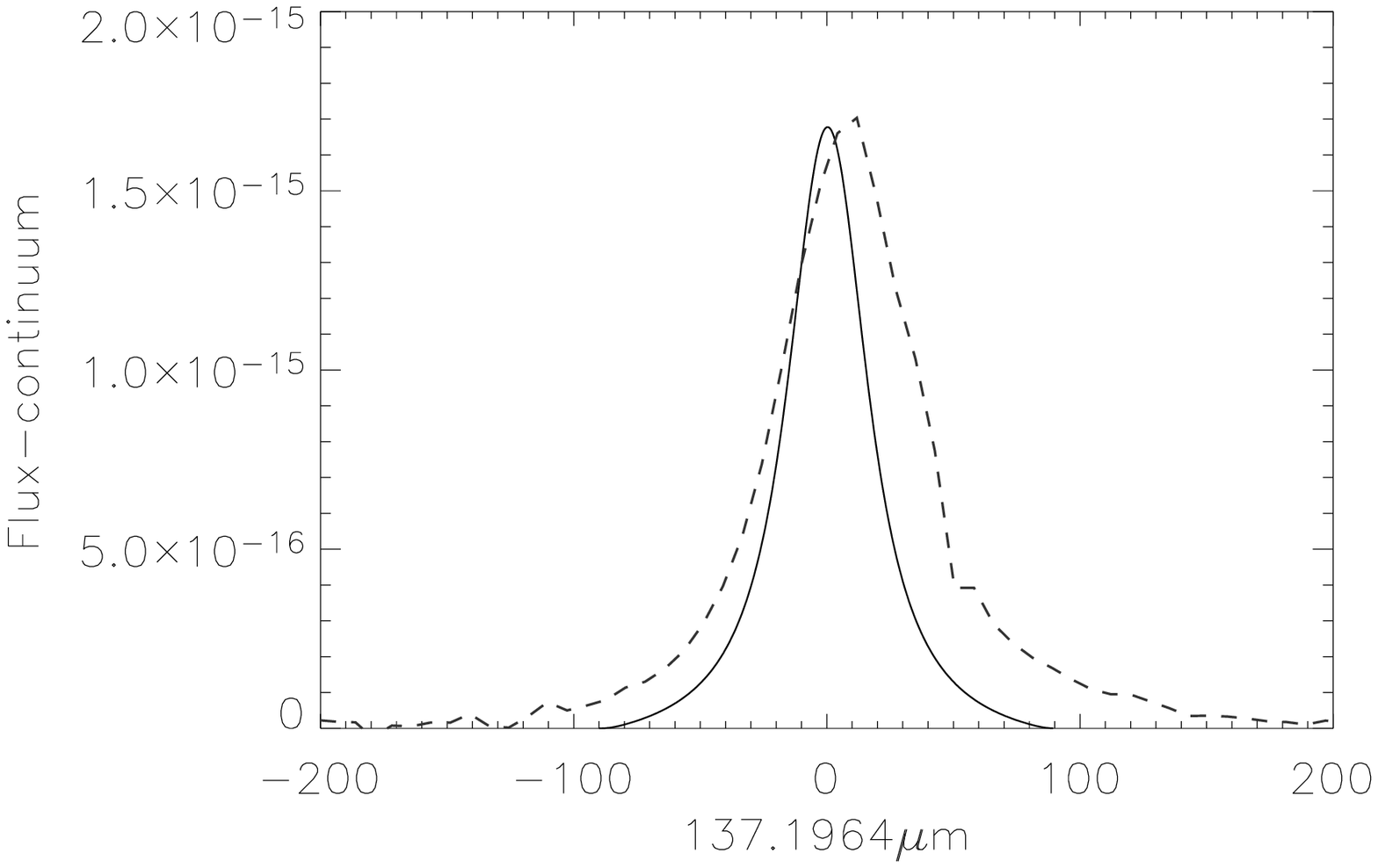}
\includegraphics[width=4cm,height=4cm]{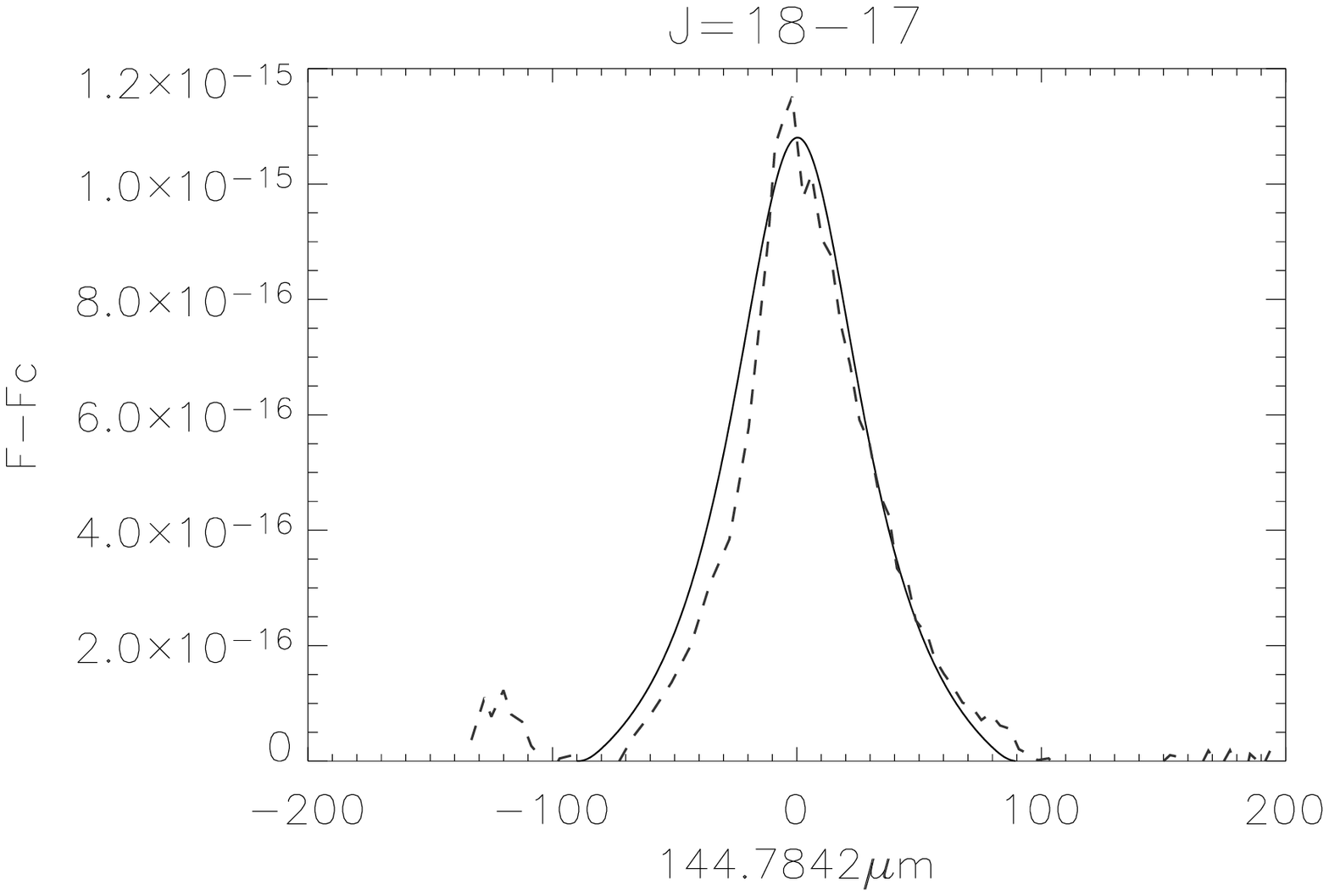}
\includegraphics[width=4cm,height=4cm]{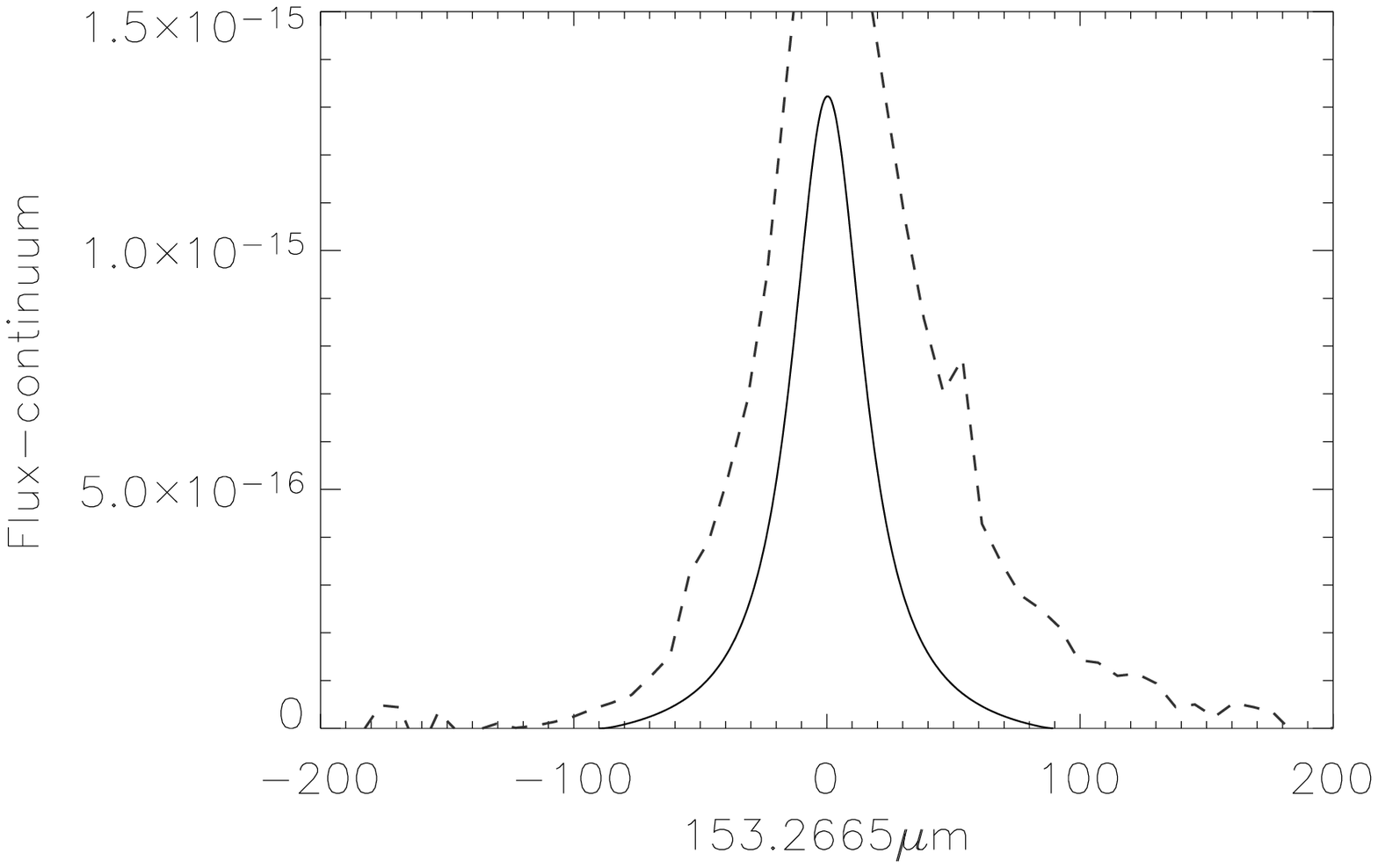}
\includegraphics[width=4cm,height=4cm]{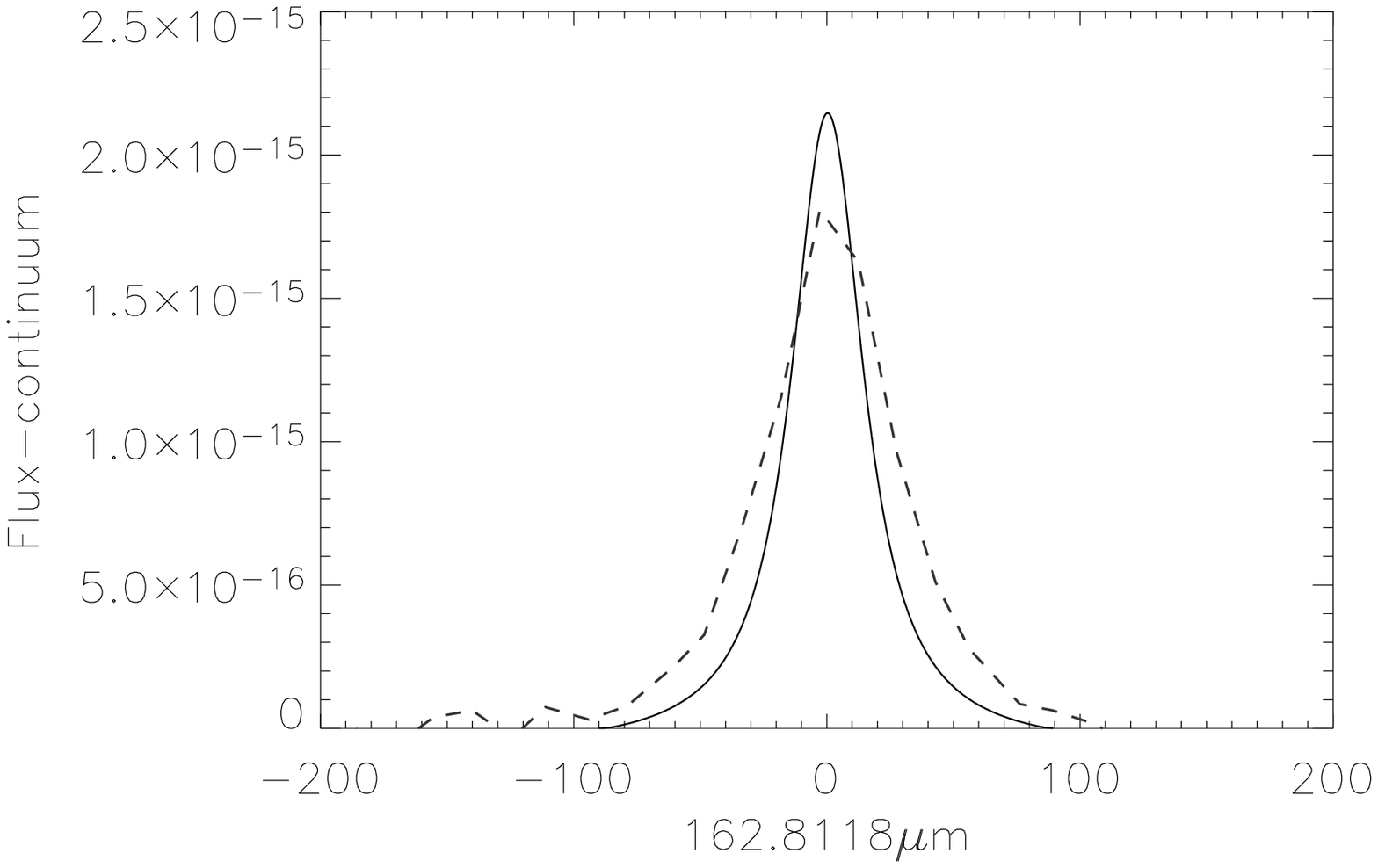}
 \caption{Continued}
 \label{co_results}
  \end{figure}


\section{Discussion}
\subsection{The Extended warm component}
Our model results indicate that the lowest energy far-IR CO
transitions originate from warm ($\approx$ 300 K) and dense
($\approx$ 10$^{5}$ cm$^{-3}$) gas. The derived temperature and
density agree reasonably well with those from the combined flux
ratios of [O~{\sc i}] 63.2/145.5 $\mu$m and [O~{\sc i}]
63.2/[C~{\sc ii}] 157.7 $\mu$m which yielded a temperature of 300
K and a density of $\approx$ 10$^{5}$ cm$^{-3}$ (Lerate et al. 2006). \\

The results agree with the gas density component described in
Sempere et al. (2000) as the extended ridge emission with density
of $\approx$ 10$^{5}$ cm$^{-3}$ and temperatures $>$ 120 K.
Similar results were found by Maret et al. (2001), however they
explained the emission as due to the non-ionising ultraviolet (UV)
radiation from the
Trapezium stars, i. e., the photodissociation region (PDR). \\

\subsection{The Plateau}

The CO transitions with 18$<$J$_{up}<$30 can be fitted by a
two-component emission region.  The low-velocity plateau is
modelled by simulating a non dissociative C-type shock of
$\approx$ 30 km s$^{-1}$ which leads to an increase in gas
temperature to 1000 K in a period of $\approx$ 100 yr, reaching
its maximum temperature at an age of 1000 years,  followed by fast
cooling to temperatures of 100K (as described by Bergin et al.
1998). The density and diameter of this region are 10$^{6}$ cm $^{-3}$
and $\approx$ 0.061 pc respectively, and this model component
gives a good fit to the observed high-J CO transitions. Both
Sempere et al. (2000) and Maret et al. (2001) agree on the
presence of hot and dense gas arising from the shocked region at
the interaction of the outflows with the ambient gas. However,
their quoted temperatures (1500-2000K) and densities (up to
3$\times$10$^{7}$ cm$^{-3}$) are higher
than implied by our models.\\
The CO transitions with 30$>$J$_{up}>$25 are best fitted by a
lower temperature shock model. Here the maximum temperature is 500
K, with subsequent cooling down to 90 K. This result provides a
new finding, being the first time that these transitions are
associated with a low temperature shock.  Our interpretation is
that this emitting region may be the external plateau,
characterized by a density of 3$\times$10$^{5}$ cm$^{-3}$,
probably formed as consequence of the interaction of the many
condensations in the flow (at lower velocity than in the inner
part of the plateau) with the surrounding quiescent medium. Note
however that Sempere et al (2000) found that transitions with
18$<$J$_{up}<$33 were best reproduced by a higher density
component at 10$^{7}$ cm $^{-3}$ and T = 400 K. A similar result
is presented by Maret et al. (2001) which also includes
transitions from J$_{up}$=20--30. We found that in those intervals
only transitions with J$_{up}>$25 are emitted by the plateau, as
the remainder are better reproduced by the hot core model,
consistent with the high density required by their models.

\subsection{The Hot Core}

\begin{figure}
    \centering
   \includegraphics[width=8cm]{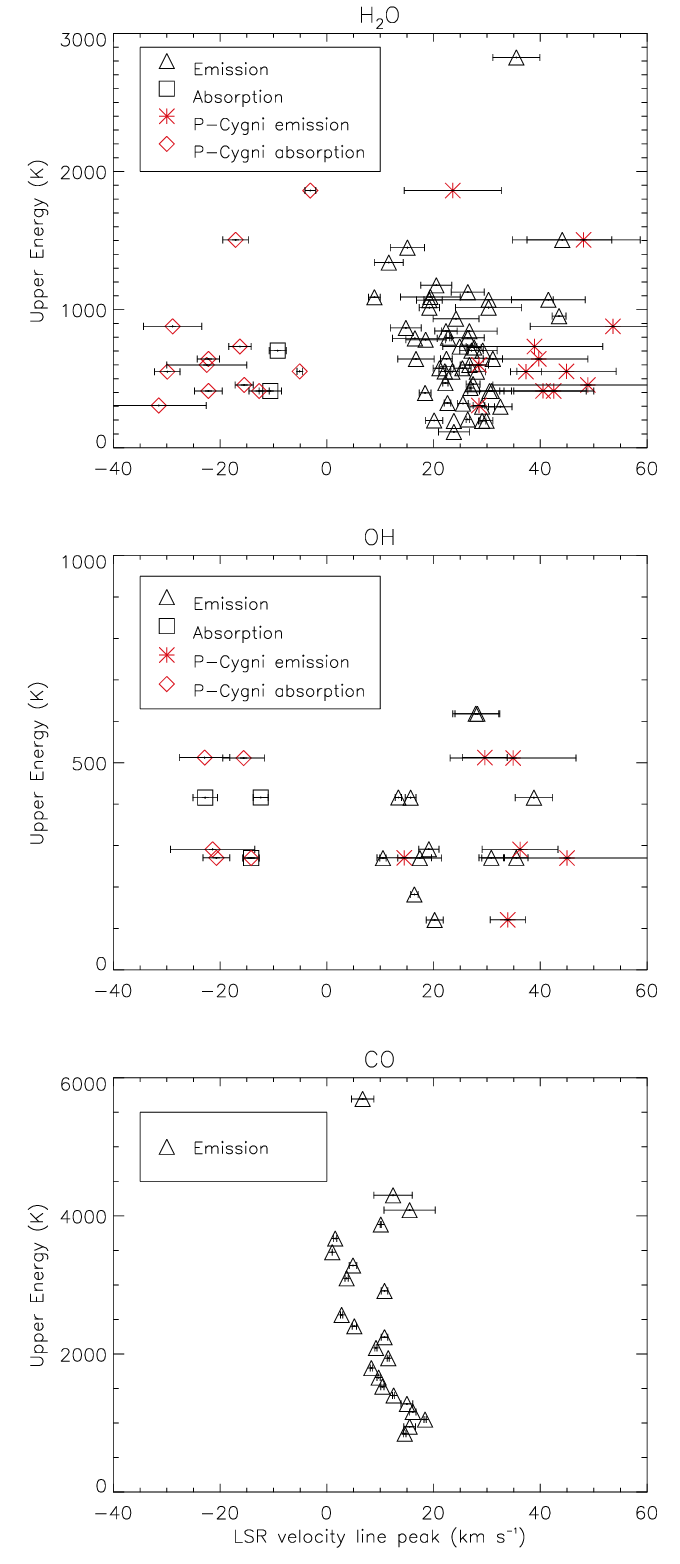}

    \caption{LSR peak velocities of H$_{2}$O, OH and CO lines observed
    towards Orion-KL with the {\em ISO} LWS-FPs (Lerate et al. 2006). Note that
    only statistical errors are plotted and do not include the overall calibration uncertainty of $\pm$ 5 km s$^{-1}$}.
        \label{velo}
\end{figure}
The Orion hot core is well known to be warm (T $\approx$ 200 K),
compact (angular size $\approx$ 10 arcsec) and dense (10$^{7}$
cm$^{-3}$) (Blake et al. 1986, Wright et al. 1992). Such a hot
core can reproduce well the CO fluxes and line profiles with
J$_{up}$=18--25. As mentioned in the plateau discussion, previous
analyses of these transitions (Sempere et al. 2000; Maret et al.
2001) explain them as coming from the plateau where a high density
and temperature $\approx$ 350--400 K component is needed to
reproduce the lines. We find that this high density component is
in fact the hot core, where the temperature is close to 200 K.
Plateau models of the same density and temperature fail to
reproduce
these transitions.\\

\subsection{Velocity structure and the far-IR H$_{2}$O and OH lines}

An analysis of the observed line radial velocities and their FWHMs
can be used to distinguish emission originating from different
components. However, the lack of spatial resolution complicates
the analysis as the dynamical properties are diluted by the large
LWS beam. The CO lines mainly peak in the LSR velocity range
between +5--15 km s$^{-1}$ (see Figure~\ref{velo}); but, because
the line centre uncertainty is $\approx$ $\pm$ 5 km s$^{-1}$ it is
difficult to establish whether lines originate from the plateau or
from the hot core component. Nevertheless, it seems clear that CO
has a different spatial origin when compared to H$_{2}$O and OH
molecules which have emission lines peaking in the range +20--40
km s$^{-1}$ LSR and lines with pure absorption profiles, which
peak at --(10--20) km s$^{-1}$. Both the OH and H$_{2}$O emission
have been modelled (Goicoechea et al. 2006, Cernicharo et al.
2006) and it was inferred that they mainly arise from an extended
flow of expanding gas at a velocity of $\approx$ 25 km s$^{-1}$
with a density of $\approx$ 5 $\times$ 10$^{5}$ cm$^{-3}$ and gas
kinetic temperature of $\approx$ 100 K. The inferred beam-averaged
abundances are $X$(H$_{2}$O)=(2--3)$\times$10$^{-5}$ and
$X$(OH)=(0.5--1)$\times$10$^{-6}$ (Cernicharo et al. 2006,
Goicoechea et al. 2006). These physical conditions are close to
those found by our model of the plateau; however, we find that the
water and OH abundances predicted by our chemical models are
respectively higher and lower than those derived by Cernicharo et
al and by Goicoechea et al. from radiative transfer model of the
observed lines (see
Figure~\ref{PL1}).\\
For our models, the higher water abundance is due to (i) high
efficiency in the hydrogenation of oxygen on the grains in Stage
I, followed by evaporation of the icy mantles in Stage II; and
(ii) high temperature reactions occurring during the shock phase
(Elitzur and Watson 1978). The latter consist of:
\begin{equation}
O + H_{2} \rightarrow OH + H
\end{equation}
\begin{equation}
 OH + H_{2} \rightarrow H_{2}O + H
\end{equation}

 At higher temperatures, OH is also formed via molecular oxygen
destruction:
\begin{equation}
O_{2} + H_{2} \rightarrow OH + OH
\end{equation}
 The high temperatures produced by the shock are necessary to
explain the OH abundance of $X$(OH)=(0.5--1)$\times$10$^{-6}$. A
shock temperature of at least 1000 K is required to reproduce the
OH abundance.
Possible explanations for the differences between the physical
conditions inferred for the CO-emitting regions and those inferred
for the H$_2$O- and OH-emitting regions include:
\begin{itemize} \item The latter two molecules arise from another
physically
distinct component, with no contribution from the hot core or the inner
plateau. \item The inferred H$_{2}$O and OH
molecular abundances are in fact very diluted by the large LWS beam,
where the contribution from shocked regions is not resolved. In
this case, the use of a radiative transfer model alone is not enough
to distinguish between the different components, and the resulting
abundances may be locally underestimated.
\end{itemize}

\section{conclusions}

We have analysed the final calibrated high resolution ISO LWS-FP
observations of CO lines using chemical models coupled with a
non-local Accelerated Lambda Iteration (ALI) radiative transfer
model. The main physical parameters of the different components
(density, temperature, scale, interstellar radiation field, dust
temperature) are introduced in the chemical models together with
physical parameters based on literature values for the hot core,
plateau, ridge and the PDR surrounding the KL region. This
technique allows a better approach to the difficult task of
distinguishing between the emission from the different components
included in the large ISO-LWS beam. Our modelling results show
that CO transitions with 18$<$J$_{up}<$25 are reproduced by a hot
core model of density $\approx$ 10$^{7}$ cm$^{-3}$ and a
temperature of 200 K. This result suppose a new interpretation of
the CO lines origin, as these transitions have been previously
interpretated as originated in the Plateau component. Transitions
between 25$<$J$_{up}<$39 are reproduced by plateau models with
densities
 ranging between 3$\times$10$^{5}$--1$\times$10$^{6}$ cm$^{-3}$ simulating
 shocked material at 500--1000 K. Derived CO abundances
 are summarised in Table~\ref{abundances}.\\
  Our chemical models do not agree with the abundances previously found from the far-IR H$_{2}$O and
  OH emission lines (Cernicharo et al., 2006; Goicoechea et al.,
  2006). Therefore, we conclude  that these species originate in a physically different
region to that responsible for the CO emission. It is also
possible that the inferred H$_{2}$O and OH
   abundance are underestimated due to the enhanced contributions
   from shocked components being diluted in the large LWS beam.
   This is because H$_{2}$O and
  OH emission lines were modelled by a different technique by the
  use of only a radiative transfer model, in which only one abundance for
  both species needed to be assumed.

We found the technique used here to be very efficient for understanding
the chemistry and ideal for modelling high spectral and spatial
resolution data.  Future heterodyne instruments, such as HIFI on the
ESA Herschel Space Observatory, with their much higher spatial and
spectral resolution, will be better able to distinguish the
different components in the complex region of Orion-KL and,
together with modelling techniques such as those developed here, should
be capable of providing new insights into cloud chemical evolution during
the process of massive star formation.

\section*{Acknowledgements}
      This work was carried out on the Miracle Supercomputer, at the HiPerSPACE Computing Centre,
      UCL, which was funded by the U.K. Particle Physics and Astronomy
Research Council.
      The {\em ISO} Spectral Analysis Package (ISAP) is a joint
      development by the LWS and SWS Instrument Teams and Data
      Centres. Contributing institutes are CESR, IAS, IPAC, MPE,
      RAL and SRON. LIA is a joint development of the {\em ISO}-LWS
      Instrument Team at Rutherford Appleton Laboratories (RAL,
      UK- the PI institute) and the Infrared Processing and
      Analysis Center (IPAC/Caltech, USA).\\
      SV acknowledges financial support from an individual PPARC
      Advance Fellowship.\\
      JRG was supported by a Marie Curie Intra-European Fellowship
      under contract MEIF-CT-2005-515340 within the 6th European Community Framework programme.
We are grateful for the comments of an anonymous referee.

\end{document}